\documentclass[12pt]{article}
\usepackage[utf8]{inputenc}

\usepackage{amsmath}
\usepackage{amssymb}
\usepackage{mathtools}
\usepackage{braket}
\usepackage{tikz}
\usetikzlibrary{patterns}
\usepackage{pgfplots}
\pgfplotsset{compat=1.17}
\usepackage{graphicx}
\usepackage{color}
\usepackage{cancel}
\usepackage{tensor}
\usepackage{stackrel}
\usepackage{tikz-cd}
\usepackage{cite}
\usepackage{tabularx}
\usepackage{authblk}
\usepackage{float}
\usepackage[margin=0.75in]{geometry}
\usepackage{hyperref}

\DeclareMathOperator{\sinc}{sinc}

\begin{document}

\title{Lorentz-covariant sampling theory for fields}

\author{Jason Pye\footnote{\href{mailto:jason.pye@uwa.edu.au}{jason.pye@uwa.edu.au}}}
\affil{\normalsize Department of Applied Mathematics, University of Waterloo, Waterloo, ON N2L 3G1, Canada}
\affil{\normalsize Department of Physics, The University of Western Australia, Perth, WA 6009, Australia}

\date{\normalsize 16 January 2023}

\maketitle

\begin{abstract}
Sampling theory is a discipline in communications engineering involved with the exact reconstruction of continuous signals from discrete sets of sample points. From a physics perspective, this is interesting in relation to the question of whether spacetime is continuous or discrete at the Planck scale, since in sampling theory we have functions which can be viewed as equivalently residing on a continuous or discrete space. Further, it is possible to formulate analogues of sampling which yield discreteness without disturbing underlying spacetime symmetries. In particular, there is a proposal for how this can be adapted for Minkowski spacetime. Here we will provide a detailed examination of the extension of sampling theory to this context. We will also discuss generally how spacetime symmetries manifest themselves in sampling theory, which at the surface seems in conflict with the fact that the discreteness of the sampling is not manifestly covariant. Specifically, we will show how the symmetry of a function space with a sampling property is equivalent to the existence of a family of possible sampling lattices related by the symmetry transformations.
\end{abstract}

\section{Introduction}

Sampling theory is a collection of mathematical tools used extensively in communications engineering for the reconstruction of continuous signals \emph{exactly} from a discrete set of sample points.
In communication theory, one is typically dealing with functions which are bandlimited (i.e., whose Fourier transforms have bounded support).
The central result of sampling theory is the demonstration that a function, $\phi$, bandlimited to a finite interval $(-\Omega,\Omega)$ in Fourier space, can be reconstructed from its values at a discrete set of sample points:
\begin{equation}\label{eq:shannon}
  \phi(x) = \sum_{n \in \mathbb{Z}} \sinc \left[ \Omega \left( x - \frac{n\pi}{\Omega} \right) \right] \phi \left( \frac{n\pi}{\Omega} \right),
\end{equation}
where $\sinc(x) := \sin(x)/x$ (and equal to $1$ at $x=0$).
This is referred to as the Shannon sampling theorem, stemming from Shannon's application of the formula in communication theory \cite{Shannon1948,Shannon1949}.
There are other names also associated with this formula, corresponding to the earlier works of Whittaker \cite{Whittaker1915}, Nyquist \cite{Nyquist1928}, and Kotel'nikov \cite{Kotelnikov1933}.
The theorem shows that we are free to choose whether we represent a bandlimited function by the continuous signal, $\phi(x)$, or the discrete samples, $\phi \left( \frac{n\pi}{\Omega} \right)$, since the reconstruction formula \eqref{eq:shannon} establishes that they contain equivalent information.

Since the time of Shannon, information theory has been significantly expanded to account for the quantum nature of physical systems, with applications ranging from quantum technologies to black hole thermodynamics.
There is also a growing body of literature investigating the transmission of information through quantum fields in the relativistic regime \cite{PeresTerno2004,BradlerHaydenPanangaden2009,ClicheKempf2010,Kent2012,BradlerHaydenPanangaden2012,RideoutEtal2012,DownesRalphWalk2013,MartinmartinezAasenKempf2013,BruschiEtal2014,Martinmartinez2015,Landulfo2016,Jonsson2017,JonssonRiedMartinmartinezKempf2018,SimidzijaAhmadzadeganKempfMartinmartinez2020}.
It is then natural to extend sampling theory, which is so important in classical information theory, to quantum fields.
It has also been suggested that a fundamental minimum length at the Planck-scale may effectively yield some form of bandlimitation \cite{Kempf2000,Kempf2004b}.
This leads one to consider the role of such a bandlimit, and the corresponding sampling theory, for the fundamental information theory of quantum fields.

The possibility that there may be a fundamental limitation to the scale at which the quantized gravitational field can be probed was first identified by Bronstein in \cite{Bronstein1936,Bronstein2012}.
Further thought experiments suggest that limitations to the precision of distance measurements may be a general feature of physics at the Planck scale ($\ell_P := \sqrt{\hbar G / c^3} \sim 10^{-35}m$), as a result of combining the uncertainty principle for matter with the gravitational field equations \cite{Mead1964}.
This would be an unavoidable consequence of the universal coupling to gravity, and therefore might be thought of as a fundamental property of spacetime.

A minimal length scale is a feature of many approaches to quantum gravity (see reviews \cite{Garay1995,Hossenfelder2013}).
For example, in loop quantum gravity, this appears as a finite minimum eigenvalue of the area and volume operators (see, e.g., \cite{Rovelli2004,RovelliVidotto2015}).
In string theory, one can show that there is a limit to the spatial resolution that can be achieved with strings, which can be expressed in the form of a modified uncertainty principle (see, e.g., \cite{AmatiCiafaloniVeneziano1989,KonishiPaffutiProvero1990,Witten1996,Yoneya2000}).
Instead of first constructing a theory of quantum gravity, one could attempt the more modest task of developing some notion of minimal length that can be introduced into quantum field theory and the Standard Model, in order to get a picture of how they may be modified as one approaches the Planck scale from lower energies.
For example, it has long been speculated that Planck-scale physics will provide a regulator for divergences which occur in quantum field theory \cite{Deser1957,DeWitt1964}.
One divergent quantity of interest is the entanglement entropy between local subsystems in quantum fields.
This quantity plays a key role in black hole thermodynamics and the holographic principle \cite{Bekenstein1973,Hawking1975,Sorkin1983,BombelliKoulLeeSorkin1986,Srednicki1993,SusskindLindesay2005,JacobsonWall2010,Solodukhin2011,AlmheiriMarolfPolchinskiSully2013,AlmheiriEngelhardtMarolfMaxfield2019,Penington2020,AlmheiriEtal2021}.
There are also various ideas about how the structure of spacetime might emerge from quantum entanglement \cite{Jacobson1995,VanRaamsdonk2010,BianchiMyers2014,Jacobson2016,CaoCarrollMichalakis2017,Hoehn2017}.
In these works, it is often assumed that the entanglement entropy is finite, and it may be useful for these studies to understand more precisely the manner in which it might be regulated.

Of course, eventually one would like to match such modifications of quantum field theory with low-energy limits of quantum gravity theories.
However, studying a range of possible lower energy modifications may also yield tests for certain aspects of such a minimum length structure (see, e.g., \cite{Amelinocamelia2002a,Hossenfelder2011,LiberatiMaccione2011}).
One example is to look for signatures of Planck-scale physics in the cosmic microwave background (CMB) \cite{Jacobson1999,MartinBrandenberger2001,BrandenbergerMartin2001,BrandenbergerMartin2002,Kempf2001,Niemeyer2001,KempfNiemeyer2001,EastherEtal2001,EastherEtal2002,EastherEtal2003,Shiu2005}.
In inflationary models, inhomogeneities in the CMB are due to quantum field fluctuations only a few orders of magnitude away from the Planck scale.
It is therefore possible that one could observe some impact that a minimal length had on these fluctuations.

The problem with a naive model for spacetime with a minimal length scale, such as simply replacing spacetime with a discrete lattice, is that it breaks Lorentz symmetry, which is crucial to the structure of the Standard Model.
There is currently no evidence to suggest that Lorentz symmetry breaks down \cite{Mattingly2005,KosteleckyRussell2011,Liberati2013,HAWC2020,MAGIC2020} (although there are still only few probes of spacetime at the Planck scale).
Some models for spacetime suggest that the continuum is modified, but the usual Lorentz symmetry of the spacetime is only altered in subtle ways.
For example, there are models with a curved momentum space where Lorentz symmetry is not broken but becomes deformed as one approaches the Planck scale \cite{MagueijoSmolin2002,Amelinocamelia2002b,AmelinocameliaFreidelKowalskiglikmanSmolin2011,Snyder1947}.
Another example is causal set theory, where spacetime is discrete, but in a way that does not pick out a preferred Lorentz frame \cite{BombelliLeeMeyerSorkin1987,BombelliHensonSorkin2009,Dowker2011,DowkerSorkin2020}.

Here we will examine yet another possibility, inspired by sampling theory, for introducing a minimal length to quantum field theory, which we view as perhaps complementary to these other models.
Indeed, as we will explain in the next section, this sampling-theoretic approach to modelling Planck-scale discreteness is closely related to Generalized Uncertainty Principles (GUPs) \cite{Maggiore1993a,Maggiore1993b,Maggiore1994,Kempf1994,KempfManganoMann1995,Kempf2000}.
The fact that bandlimited functions contain equivalent continuous and discrete representations is compelling when considered in relation to the tension between spacetime symmetries and a minimal length at the Planck scale.
Intuitively, continuity of spacetime would seem to be necessary to preserve spacetime symmetries, yet a fundamental minimum length would suggest some form of discreteness.
Sampling theory shows how, in a sense, we can have both.
The idea of this approach to modelling a fundamental minimum length is then to consider fields on spacetime to exhibit some form of bandlimitation, so that they possess equivalent continuous and discrete representations \cite{Kempf2000,Kempf2004b}.
So far, we have only discussed sampling in one dimension, but in \cite{Kempf2004a} it was shown how certain aspects can be generalized to curved spacetimes.
It has also been considered how these ideas could be applied to spacetime itself \cite{Kempf2009,Kempf2010,Kempf2018}.
We will elaborate upon generalized forms of bandlimitation further in the paper, with a particular focus on the case of Minkowski spacetime.

Given some appropriate notion of bandlimitation, it can be applied to quantum field theory in its path integral formulation by restricting the integration to the space of bandlimited fields.
In \cite{KempfChatwindaviesMartin2013}, it was shown how the covariant notion of bandlimitation on Minkowski spacetime modifies the propagators of scalar quantum fields.
This was then extended to inflationary spacetimes to show the impact of the bandlimit on the CMB spectrum \cite{KempfChatwindaviesMartin2013,ChatwindaviesKempfMartin2017}.
A Euclidean version of bandlimitation in quantum field theory was also studied in \cite{PyeDonnellyKempf2015,HendersonMenicucci2020} to see its effect on entanglement.

In this paper, our focus will be on developing and clarifying some of the mathematics and general features of sampling theory in Minkowski spacetime.
In particular, we are first going to be examining in detail the analogue of the reconstruction formula \eqref{eq:shannon} for Minkowski spacetime.
This was stated previously in \cite{KempfChatwindaviesMartin2013}, but here we provide a more complete derivation.
Elaborating upon the steps in the development of the formula will help us identify certain features of the reconstruction that were not previously recognized, and perhaps could be useful in future applications.
We will then proceed to clarify how the discreteness of the sampling theory is consistent with the continuous spacetime symmetries.
Indeed, the fact that there is an equivalent continuous representation is not sufficient to establish that there is a symmetry (for example, a simple interpolation of a discrete set of points does not necessarily yield a Lorentz-invariant function space).
Previously, some arguments were made as to how one can think of this in the Minkowski case, but we will explain how the situation is more subtle.
In the one-dimensional scenario the manifestation of the symmetry in the sampling theory is well understood: it appears as a family of translated sampling lattices that can each be used for the sampling.
We will show on general grounds how this idea can be extended to any function space with a symmetry and a sampling property.

The outline of the paper is as follows.
First, in Section~\ref{sec:gups}, we will elaborate upon the connection between GUPs and sampling theory.
In Section~\ref{sec:shannon_sampling}, we will give an overview of some concepts in sampling theory which are relevant to the physical motivations given above.
In Section~\ref{sec:minkowski_sampling}, we will review the analogue of bandlimitation on Minkowski spacetime, as well as provide a detailed development of the corresponding reconstruction formula.
Finally, in Section~\ref{sec:sampling_symmetries}, we will discuss how symmetries manifest themselves in sampling theory.

\section{Connection with Generalized Uncertainty Principles}
\label{sec:gups}

A simplified version of the general argument in \cite{Bronstein1936,Bronstein2012,Mead1964} is that, in the presence of gravity, at some high-energy scale an increasing uncertainty in momentum should begin to yield an increasing uncertainty in the geometry of spacetime, which in turn causes an increased uncertainty in position or distance measurements.
Increasing $\Delta X$ with increasing $\Delta P$ is in stark contrast to the ordinary quantum mechanical relationship $\Delta X \Delta P \geq \hbar/2$.
The idea of Generalized Uncertainty Principles (GUPs) \cite{Maggiore1993a,Maggiore1993b,Maggiore1994,Kempf1994,KempfManganoMann1995,Kempf2000} is to combine the usual uncertainty principle with a modification capturing the anticipated high-energy behaviour due to gravity, where $\Delta X$ increases with increasing $\Delta P$.
A simple GUP achieving this is:
\begin{equation}
  \Delta X \Delta P \geq \frac12 ( 1 + \beta \Delta P^2 ).
\label{eq:gup}
\end{equation}
Including this additional term on the right-hand side implies $\Delta X \geq \sqrt{\beta}$ (assuming $\beta > 0$).
Hereafter we work in Planck units, where $\hbar = c = G = 1$.
The parameter $\sqrt{\beta}$ is typically assumed to be on the order of the Planck length, which is $1$ in these units.

A modified uncertainty relation of this kind can be modelled by a change in the commutation relations between $X$ and $P$ \cite{Maggiore1994,Kempf1994,KempfManganoMann1995}, for example,
\begin{equation}
  [ X, P ] = i ( 1 + \beta P^2 ).
\label{eq:gup_commutator}
\end{equation}
This would then immediately yield the above GUP:
\begin{equation}
  \Delta X \Delta P \geq \frac12 ( 1 + \beta \Delta P^2 + \beta \langle P \rangle^2 ) \geq \frac12 ( 1 + \beta \Delta P^2 ),
\end{equation}
assuming $\beta > 0$.
There have been many extensions of this idea, as well as applications to various physical scenarios, see, e.g., \cite{Kempf1997,Brau1999,HossenfelderEtal2003,DasVagenas2008,PikovskiEtal2012,ScardigliCasadio2015,ScardigliBlasoneLucianoCadadio2018,TodorinovBossoDas2019}.

Recently there have been proposals in the GUP literature about how one can obtain a finite minimum $\Delta X$ by modifying the position and momentum operators in a way that retains the canonical commutation relations \cite{BishopEtal2020,BishopEtal2021,BishopEtal2022a,BishopEtal2022b} (see also \cite{BossoEtal2022}).
One example from \cite{BishopEtal2020,BishopEtal2021,BishopEtal2022a,BishopEtal2022b} is to define (in slightly different notation),
\begin{equation}
  (X \psi)(p) := i ( 1 + \beta p^2 ) \frac{d}{dp} \psi(p), \qquad (K \psi)(p) := \frac{1}{\sqrt{\beta}} \text{arctan} ( \sqrt{\beta} p ) \psi(p),
\label{eq:xk_reps}
\end{equation}
where $\psi(p)$ is a square-integrable function with respect to the inner product $\langle \phi | \psi \rangle = \int \tfrac{dp}{2\pi} (1+\beta p^2)^{-1} \phi^\ast(p) \psi(p)$.
These two operators satisfy $[X,K] = i$.
As opposed to the GUP \eqref{eq:gup}, where a finite minimum $\Delta X$ is implied by the abstract algebraic relation, here it is implied by $\Delta X \Delta K \geq 1/2$ and the fact that the spectrum of $K$ is bounded on this domain (hence there is an upper bound to $\Delta K$, and thus a lower bound for $\Delta X$).
Note that here we can also reintroduce a second momentum operator $P$ as $(P\psi)(p) = p\psi(p)$, which again satisfies $[X,P] = i(1+\beta P^2)$ and implies a finite minimum $\Delta X$ in the same way as before.
Therefore, the difference between the original modified commutator \eqref{eq:gup_commutator} and the proposal \eqref{eq:xk_reps} is in the identification of the physical momentum operator.
Both pairings can be represented on the same function space, and imply the same finite minimum $\Delta X$ (though in different ways).

We note here that changing coordinates in momentum space by,
\begin{equation}
  k = \frac{1}{\sqrt{\beta}} \text{arctan}(\sqrt{\beta}p),
\end{equation}
(as done in \cite{Pye2019}, for example) yields an equivalent $K$-representation for the above three operators:
\begin{equation}
  (X\psi)(k) = i \frac{d}{dk}\psi(k), \qquad (K\psi)(k) = k \psi(k), \qquad (P\psi)(k) = \frac{1}{\sqrt{\beta}} \text{tan} ( \sqrt{\beta} k ) \psi(k),
\end{equation}
where $\psi(k)$ is square-integrable with respect to the inner product $\langle \phi | \psi \rangle = \int_{-\pi/2\sqrt{\beta}}^{\pi/2\sqrt{\beta}} \tfrac{dk}{2\pi} \phi^\ast(k) \psi(k)$.
Because of the finite range of $\text{arctan}$, functions in the $K$-representation are only supported on the interval $(-\frac{\pi}{2\sqrt{\beta}},\frac{\pi}{2\sqrt{\beta}})$, i.e., they are bandlimited by $\Omega = \frac{\pi}{2\sqrt{\beta}}$.
Also, $X$-space and $K$-space are simply related by the usual Fourier transform.
Therefore, the space of wavefunctions one obtains in $X$-space is the same space of functions that is studied in Shannon sampling theory.
We then see that the space of bandlimited functions is simply a different representation of the space of functions studied with \eqref{eq:gup} and \eqref{eq:xk_reps}.

In \cite{Pye2019}, it was shown how one can analogously relate certain Lorentz-covariant versions of the GUP to bandlimitation on Minkowski spacetime.
One begins with a set of modified commutators of the form:
\begin{equation}
  [ X_I, P^\mu ] = i \tensor{\theta}{_I^\mu}(P), \qquad [ X_I, X_J ] = 0, \qquad [ P^\mu, P^\nu ] = 0.
\end{equation}
Note that the coordinates remain commuting.
With certain technical assumptions on $\tensor{\theta}{_I^\mu}$ (see \cite{Pye2019} for details), we can introduce operators $K^I$ which are functions of the operators $P^\mu$ and satisfy canonical commutation relations $[ X_I, K^J ] = i \tensor{\delta}{_I^J}$.
Similar to before, the operators $X_I$, $P^\mu$, and $K^J$ can be represented equivalently in either the $P$- or $K$-representations, which are simply related by a change of momentum-space coordinates.
The functions $\tensor{\theta}{_I^\mu}$ can be chosen to yield a cutoff in $K$-space, similar to the one-dimensional case.
For example, if the two momentum spaces are related by $p^I = k^I/[1 - (k^2/\Omega^2)^2]$, where $k^2 := k_0^2 - \vec{k}^2$, then functions in $K$-space are bandlimited to $|k^2| < \Omega^2$.
Note that since $X_I$ and $K^J$ obey canonical commutation relations, infinitesimal Lorentz transformations can be represented by using the generators $L_{IJ} := X_I K_J - X_J K_I$.
In the case where we have a Lorentz-invariant cutoff $|k^2| < \Omega^2$, the corresponding function space will then also be invariant under finite Lorentz transformations.

Above we said that $X$-space and $K$-space are related by a Fourier transform because these operators obey canonical commutation relations.
However, even in the one-dimensional case, this is perhaps unclear due to the fact that a finite minimum $\Delta X$ means that $X$ has no eigenvectors (nor sequences of states yielding approximate eigenvectors) in the domain on which the commutation relations hold \cite{Kempf1994,KempfManganoMann1995}.
The functional analysis of the position operator in the context of modified commutation relations (or equivalently the bandlimitation scenario) was discussed thoroughly in \cite{KempfManganoMann1995,Kempf2000,Kempf2004b}.
There it was found that there is a one-parameter family of self-adjoint extensions of the position operator.
Each operator in this family provides a lattice on which bandlimited functions can be represented.
The union of all of these lattices covers $\mathbb{R}$ once, and the coefficients of states in the corresponding overcomplete basis can be interpreted as a bandlimited function in a position space which is simply Fourier-related to the $K$-representation.
These functions can therefore be represented equivalently in the overcomplete continuous basis, or any one of the lattices corresponding to a particular self-adjoint extension.
In \cite{Kempf2000,Kempf2004b} it was recognized that this was describing the mathematics behind Shannon sampling theory.
Note that these observations also apply to position operators of \eqref{eq:gup} and \eqref{eq:xk_reps}, since we established that these describe the same function space (and the same position operator), but with different momentum-space coordinates (and possibly a different physical momentum operator).
Sampling theory therefore provides a general framework for studying the position-space representations of position operators which have a minimum uncertainty.
These observations further motivated a general mathematical correspondence between operators with finite minimal uncertainty and function spaces which exhibit a sampling theorem \cite{Kempf2000,Kempf2004b,MartinKempf2009,Martin2010,MartinKempf2018}.
Also developed from this was a notion of time-varying bandwidth, which has led to interesting applications in engineering \cite{HaoKempf2007,HaoKempf2008,HaoKempf2010a,HaoKempf2010b} as well as connections to pure mathematics \cite{Martin2011,Martin2015,PunMartinKempf2018}.

We note also recent work in \cite{LakeEtal2019,LakeEtal2020}, wherein a model for a minimal length is obtained as a consequence of including additional quantum degrees of freedom to implement a smearing of spatial points.
In these works, a position operator is constructed which exhibits a finite minimum uncertainty, but acts on a space of functions that does not appear to be bandlimited.
The general result of \cite{MartinKempf2009} would suggest that the finite minimum uncertainty should yield a corresponding sampling property.
A careful examination of the functional analysis of this position operator would be required to clarify the connection to sampling theory.
Such an undertaking would be too lengthy to include here, but it would be interesting since it may demonstrate a sampling theorem which does not arise from bandlimitation.

We end this section by noting that, for one-dimensional Shannon sampling, translation symmetry of the bandlimited function space is reflected in the fact that none of the family of lattice representations are preferred.
As we will demonstrate in Section~\ref{sec:minkowski_sampling}, one can construct an analogous sampling property in the case of a Lorentz-invariant bandlimit $|k^2| < \Omega^2$.
However, carrying out the analogous functional-analytic arguments to connect Lorentz-symmetry with a family of lattice representations is not as straightforward.
For our purposes here, we will not be concerned with the functional analysis of the position operators.
Instead, we will simply \emph{define} the position-space representation of a function as the Fourier transform of its $K$-representation.
Our interest will then be in sampling formulas for these functions in position space.
In Section~\ref{sec:sampling_symmetries}, we will then show on more general grounds how one can reconcile Lorentz symmetry with the sampling representations.

\section{Sampling theory}
\label{sec:shannon_sampling}

In this section we will review some basic concepts in sampling theory, with an emphasis on those which may be of interest in physical applications.
We will also review a simple derivation of the Shannon sampling formula.
An extension of the method employed in this derivation will be used in the next section to develop an analogous sampling formula for bandlimited functions on Minkowski spacetime.

\subsection{Overview}
\label{subsec:sampling_overview}

The typical focus of sampling theory is the reconstruction of some particular class of functions on a space from a discrete set of sample points.
The original sampling theorem \eqref{eq:shannon} applies to functions on $\mathbb{R}$ with frequency support in a bounded interval, but the basic idea of reconstructing functions from sample values has been extended to many other contexts in communication theory (see, e.g., \cite{Jerri1977,Zayed1993,Higgins1996,BenedettoFerreira2001}).
For example, similar results apply to bandlimited functions whose Fourier transforms have support on more general subsets of finite measure.
One can also consider analogues of sampling in higher dimensions and on manifolds different from $\mathbb{R}^N$.
Going in this direction, the first task is to determine an appropriate generalization of bandlimitation.
For example, a natural analogue of bandlimitation in higher-dimensional Euclidean space as well as Minkowski spacetime is a restriction in Fourier space to some subset of momenta, $\vec{k}$, or four-momenta, $k^\mu$, respectively.

Here we will be particularly interested in restrictions which preserve symmetries.
In the one-dimensional scenario, where we restrict $|k| < \Omega$ and impose Dirichlet boundary conditions, we obtain a bandlimited subspace which is invariant under translations (we will discuss this in more detail in Section~\ref{sec:sampling_symmetries}).
In higher-dimensional Euclidean space, a simple bandlimited subspace which is invariant under both translations and rotations consists of functions whose Fourier transforms have support in the region $\vec{k}^2 < \Omega^2$, for some fixed $\Omega > 0$.
Similarly, in Minkowski spacetime, a bandlimited subspace invariant under Poincar\'e transformations consists of wavepackets of the form:
\begin{equation}
  \phi(x) = \int_{|k^2| < \Omega^2} \frac{d^4k}{(2\pi)^4} \tilde{\phi}(k) e^{-i k \cdot x},
\end{equation}
where $k^2 \equiv k_\mu k^\mu$ and $k \cdot x \equiv k_\mu x^\mu$, with metric signature $(+,-,-,-)$.
This latter extension of the notion of bandlimitation to Minkowski spacetime was first proposed by Kempf in \cite{Kempf2004a}.
Here the momentum space cutoff is considered as a physical Lorentz-invariant cutoff.
Note that this is different from introducing a formal Euclidean regulator, $k_E^2 < \Omega^2$, after a Wick rotation.
In particular, a Euclidean cutoff will fully regulate momentum space integrals by restricting the integration region to a finite volume.
The Minkowski spacetime bandlimit generally will not.

In the settings of curved spaces and spacetimes, a generalization of plane waves are the eigenfunctions of the covariant Laplacian or d'Alembertian operator associated with a metric (namely, the operator $\Box_g = |g|^{-\frac12} \partial_\mu ( |g|^\frac12 g^{\mu \nu} \partial_\nu \cdot )$ for the metric $g_{\mu \nu}$).
Bandlimitation can then be defined as a projection onto the closure of some subset of these eigenfunctions.
In the flat space and spacetime cases, the symmetry-preserving restrictions $\vec{k}^2 < \Omega^2$ and $|k^2| < \Omega^2$ can be thought of as a projection onto the eigenspaces corresponding to the eigenvalues of $\vec{\nabla}^2$ and $\Box = \partial_t^2 - \vec{\nabla}^2$ (respectively) with magnitudes below some threshold $\Omega > 0$.
This kind of restriction can be generalized to manifolds with curvature, where one projects onto the eigenspaces of the covariant Laplacian or d'Alembertian with corresponding eigenvalues of magnitude below a cutoff $\Omega > 0$.
Because the spectra of these covariant differential operators are invariant under general coordinate transformations, so are the corresponding bandlimited subspaces.
Therefore, this kind of restriction gives us a general coordinate-invariant notion of bandlimitation.
This idea was introduced for the case of Riemannian manifolds by Pesenson in \cite{Pesenson2000,Pesenson2001} (see also developments in \cite{Kempf2004a,KempfMartin2008,MartinKempf2008}), and for pseudo-Riemannian manifolds by Kempf in \cite{Kempf2004a}.
In this paper, however, we will be primarily focused on the case of flat Minkowski spacetime.

Once a bandlimited subspace has been fixed, one aims to find an appropriate set of sample points, $\{ x_n \}_{n \in \mathbb{Z}}$, so that a bandlimited function is uniquely specified among elements in this space by its values on the sampling lattice.
Typically, one also requires the sampling to be \emph{stable}, in the sense that there is some constant $C > 0$ for which
\begin{equation}
  \| \phi \|^2 \leq C \sum_{n \in \mathbb{Z}} |\phi(x_n)|^2,
\end{equation}
where $\| \phi \|$ is the Hilbert space norm of the bandlimited function $\phi$ (we will assume that our bandlimited functions form a subspace of some Hilbert space).
The stability requirement is important in applications to ensure that noise in the signal $\phi$ does not produce arbitrarily large errors after sampling and reconstruction.
Noise present in a signal may not always be bandlimited, but typically one can ensure the signal which one samples is bandlimited by first applying an appropriate filter.

There have been important results in identifying properties of sampling lattices which are necessary and/or sufficient for the reconstruction of particular classes of bandlimited function spaces.
Often these involve concocting different notions of \emph{density} of a given sampling lattice (see Section 1.2 of \cite{BenedettoFerreira2001} for an overview).
One of the most important results is a theorem of Beurling and Malliavin \cite{BeurlingMalliavin1962,BeurlingMalliavin1967}.
They employ the following notion of density: for a given lattice $\Delta := \{ x_n \}_{n \in \mathbb{Z}} \subset \mathbb{R}$, let $\gamma$ be any interval in $\mathbb{R}$ of length $L$, and $N^-(L) := \inf_\gamma \# \{ \Delta \cap  \gamma \}$ (i.e., roughly speaking, the smallest number of samples occuring within an interval of length $L$).
Then the \emph{(lower) Beurling density} of $\Delta$ is defined as:
\begin{equation}
  D^-(\Delta) := \lim_{L \to \infty} \frac{N^-(L)}{L}.
\end{equation}
Beurling and Malliavin proved that for functions in $L^2(\mathbb{R})$ with Fourier transforms supported on a single interval, $R$, of length $|R|$, a stable sampling is possible for $\Delta$ if and only if $\Delta$ contains a uniformly discrete subset $\Delta_0$ with $D^-(\Delta_0) > |R|/2\pi$.
(Note: uniformly discrete simply means that the distance between any two distinct points is greater than some fixed positive constant.)
Therefore, functions with a larger support in Fourier space require a higher density of sample points in order to obtain a stable reconstruction.
This result also shows that the lower Beurling density, which is a kind of average density, is the only relevant feature of a lattice required to establish the possibiliy of a stable reconstruction (at least in principle).
For the case of functions bandlimited to $(-\Omega,\Omega)$, the critical density $\Omega/\pi$ was already identified in the work of Nyquist \cite{Nyquist1928}, and hence is usually referred to as the \emph{Nyquist density} in that context.

The work of Beurling and Malliavin was extended by Landau \cite{Landau1967a,Landau1967b}, providing a necessary density condition for the case of functions bandlimited to an arbitrary subset of Fourier space, as well as to the case of higher dimensions.
Specifically, Landau showed that for the space of functions in $L^2(\mathbb{R}^N)$ whose Fourier transforms are supported in a subset $R$ of Fourier space, if a stable sampling is possible for a lattice $\Delta \subset \mathbb{R}^N$, then
\begin{equation}
  D^-(\Delta) \geq \frac{|R|}{(2\pi)^N},
\end{equation}
where $|R|$ is the volume of the subset $R$ in Fourier space, and $D^-(\Delta)$ is a natural analogue of the lower Beurling density in higher dimensions (where one considers the number of sample points in $N$-dimensional balls of some fixed radius instead of intervals in the one-dimensional case, see \cite{Landau1967a} for details).
We will refer to the lower bound of $|R|/(2\pi)^N$ as the \emph{Beurling-Landau density} required for stable reconstruction in the cases considered by Landau.

Some results have also been obtained for bandlimited functions on Riemannian manifolds with curvature \cite{Pesenson2000,Pesenson2001} (see also discussion in \cite{KempfMartin2008,MartinKempf2008}).
In the pseudo-Riemannian case, the nature of the corresponding sampling is qualitatively different.
We will elaborate upon this for the Minkowski spacetime case in Section~\ref{sec:minkowski_sampling}.
An extension of the analysis of the Minkowski case to Friedmann-Robertson-Walker spacetimes was done in \cite{KempfChatwindaviesMartin2013}.
Little is known about sampling on pseudo-Riemannian manifolds beyond these examples.

The above density conditions indicate whether or not reconstruction from a given lattice may be possible in principle, but one may also wish to determine explicitly how the bandlimited functions can be reconstructed from their sample values.
Concretely, this would amount to finding an analogue of \eqref{eq:shannon}, of the general form:
\begin{equation}\label{eq:reconstruction}
  \phi(x) = \sum_{n \in \mathbb{Z}} K_n(x) \phi(x_n),
\end{equation}
for an arbitrary function, $\phi$, in the bandlimited subspace.
The function $K_n(x)$ will be referred to as a \emph{reconstruction kernel} for this bandlimited space.
In the simplest cases, $K_n$ will only depend on $n$ through the sample point $x_n$, as in \eqref{eq:shannon} with $K_n(x) = \sinc [ \Omega (x-x_n) ]$.
However, generally it may be necessary to use different kernels at different sample points (as we will see later in Subsection~\ref{subsec:Minkowski_reconstruction}).
There are a variety of techniques used for obtaining reconstruction formulas in particular scenarios (see, e.g., \cite{Jerri1977,Zayed1993,Higgins1996,BenedettoFerreira2001}).

As an illustration of some of the ideas we have presented in this section, we will now turn to a simple derivation of the Shannon sampling theorem \eqref{eq:shannon} for functions on $\mathbb{R}$ bandlimited to $|k| < \Omega$.
The method used in this derivation will develop the reconstruction kernel as well as establish the Nyquist density for this bandlimited space.
In Section~\ref{sec:minkowski_sampling}, we will then employ an extension of this method due to Kohlenberg \cite{Kohlenberg1953} to develop a kind of reconstruction formula for bandlimited functions in Minkowski spacetime.

\subsection{Derivation of the Shannon reconstruction formula}
\label{subsec:shannon_formula}

In this subsection, we will only be considering the subspace of $L^2(\mathbb{R})$ consisting of functions bandlimited to the region $|k| < \Omega$ of Fourier space.
We will first make some helpful definitions and establish some notation.
Let $\{ x_n \}_{n \in \mathbb{Z}}$ denote some discrete set of points in $\mathbb{R}$.
The \emph{sampling operator} corresponding to this set is defined as:
\begin{equation}
  S := \sum_{n \in \mathbb{Z}} |x_n)(x_n|,
\end{equation}
where $|x_n)$ denotes the position eigenvector\footnote{Note that we use round brackets $|\phi)$ instead of angled brackets $\ket{\phi}$ to denote an element of a Hilbert space so they are not confused with quantum states.
Of course, this discussion applies equally well to bandlimited wavefunctions, but here we have in mind that the Hilbert spaces physically correspond to a space of field configurations which are integrated over in a path integral.} centered at $x = x_n$.
These position eigenvectors are continuum-normalized, $(x|x') = \delta(x-x')$, form a resolution of identity, $\int dx |x)(x| = 1$, and define the position space representation of Hilbert space elements by $\phi(x) \equiv (x|\phi)$.
(Note that these position vectors have not yet been projected down to the bandlimited subspace.)
As the name suggests, the sampling operator extracts the values of a function at the discrete set of points $\{ x_n \}_{n \in \mathbb{Z}}$, by: $S |\phi) = \sum_{n \in \mathbb{Z}} \phi(x_n) |x_n)$.\footnote{Note that the sampling operator is not a well-defined operator on $L^2(\mathbb{R})$.
However, throughout this section, it will only appear in the combination $HS$, and it should be clear in what follows that with our eventual choice of a filter $H$, the operator $HS$ will be well-defined on the space of bandlimited functions.
It will simply be helpful for illustrative purposes to identify $S$ separately.}
We will also define a \emph{filter},
\begin{equation}
  H := \int \frac{dk}{2\pi} H(k) |k)(k|,
\end{equation}
as some operator on $L^2(\mathbb{R})$ which is diagonal in Fourier space.
Our momentum eigenvectors are normalized as $(k|k') = 2\pi \delta(k-k')$, and thus $\int \frac{dk}{2\pi} |k)(k| = 1$.
The Fourier-space representation of a Hilbert space element is $\tilde{\phi}(k) := (k|\phi) = \int dx \phi(x) e^{-ikx}$, where $(k|x) = e^{-ikx}$.

Now we claim that we can obtain a sampling formula if we can find a particular discrete set of sample points, $\{ x_n \}_{n \in \mathbb{Z}}$, and a filter, $H(k)$, such that the operator $HS$ is the identity on the bandlimited subspace (where $S$ is the sampling operator associated with this set of sample points).
If this can be achieved, then we immediately get a reconstruction formula of the form \eqref{eq:reconstruction} for any bandlimited function $\phi$:
\begin{align}
  \phi(x) &= (x|\phi) \\
  &= (x|HS|\phi) \\
  &= \sum_{n \in \mathbb{Z}} (x|H|x_n) \phi(x_n).
\end{align}
We can identify the reconstruction kernel as $K_n(x) = (x|H|x_n)$ in this case.
Formulating the problem in this way is not conducive to every sampling problem, but it is particularly useful when attempting to find reconstruction formulas with equidistantly-spaced sampling lattices, i.e., when $x_n = \frac{n\pi}{\Theta}$ for an appropriately chosen $\Theta$.

The simplest example is the present case of functions in $L^2(\mathbb{R})$ bandlimited to $|k| < \Omega$.
Let us consider a sampling lattice $\{ x_n = \frac{n\pi}{\Theta} \}_{n \in \mathbb{Z}}$ for some $\Theta$ to be determined.
This fixes the sampling operator, $S = \sum_{n \in \mathbb{Z}} |x_n)(x_n|$, up to the choice of $\Theta$.
Now we need to choose $\Theta$ and find a filter $H(k)$ so that $|\phi) \stackrel{!}{=} HS |\phi)$ for all $\phi$ in the bandlimited subspace.
Let us then examine this condition in Fourier space:
\begin{align}
  \tilde{\phi}(k) &\stackrel{!}{=} (k|HS|\phi) \\
  &= H(k) (k| \left[ \sum_{n \in \mathbb{Z}} |x_n)(x_n| \right] \left[ \int \frac{dk'}{2\pi} |k')(k'| \right] |\phi) \\
  &= H(k) \int \frac{dk'}{2\pi} \left[ \sum_{n \in \mathbb{Z}} e^{-i (k-k') x_n} \right] \tilde{\phi}(k') \label{eq:diraccomb1}\\
  &= H(k) \int \frac{dk'}{2\pi} \left[ 2\Theta \sum_{m \in \mathbb{Z}} \delta(k-k' - 2 \Theta m) \right] \tilde{\phi}(k') \label{eq:diraccomb2}\\
  &= \frac{\Theta}{\pi} H(k) \sum_{m \in \mathbb{Z}} \tilde{\phi}(k - 2 \Theta m).
\end{align}
The step from line \eqref{eq:diraccomb1} to \eqref{eq:diraccomb2} uses the assumption that the sample points are equidistantly spaced, so that:
\begin{equation}
  \sum_{n \in \mathbb{Z}} e^{-i (k-k') \frac{n\pi}{\Theta}} = 2\Theta \sum_{m \in \mathbb{Z}} \delta(k-k' - 2 \Theta m).
\end{equation}

\begin{figure}[htb]

  \centering
  
  \includegraphics{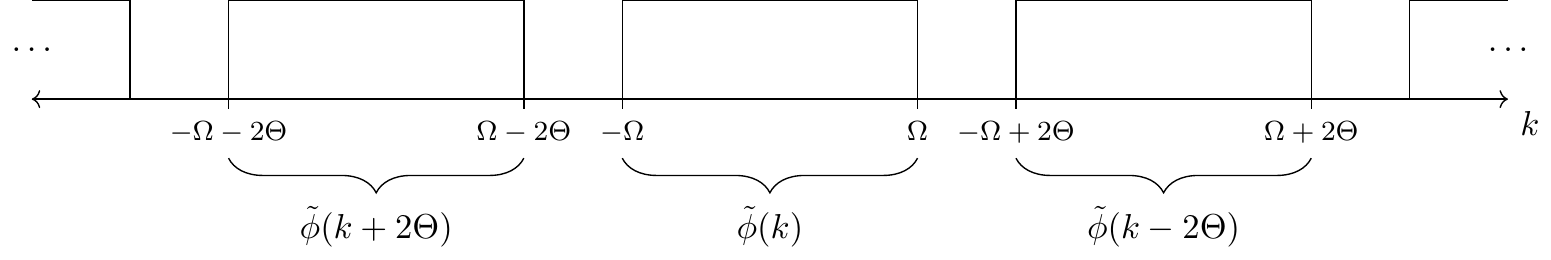}
  
  \caption{\small An illustration of the support of copies of $\tilde{\phi}(k)$ shifted by integer multiples of $2\Theta$, for $\Theta > \Omega$. The boxes are simply meant to indicate where the copies are supported; the form of $\tilde{\phi}$ can be arbitrary within this support.}
  \label{fig:copies_oversampling}
  
\end{figure}

Therefore, our sampling problem $|\phi) \stackrel{!}{=} HS |\phi)$ is to find $\Theta$ and $H(k)$ so that for every bandlimited $\phi$ we have:
\begin{equation}\label{eq:firstordercopies}
  \tilde{\phi}(k) \stackrel{!}{=} \frac{\Theta}{\pi} H(k) \sum_{m \in \mathbb{Z}} \tilde{\phi}(k - 2 \Theta m).
\end{equation}
The sum on the right-hand side of this equation is over copies of $\tilde{\phi}$ translated in Fourier space by integer multiples of $2\Theta$.
This sum can be visualized as in Figure~\ref{fig:copies_oversampling} (for $\Theta > \Omega$).
We will call the term $\tilde{\phi}(k - 2 \Theta m)$ in the sum the $m^{th}$ copy of $\tilde{\phi}$.
We see that the different copies of $\tilde{\phi}$ will be non-overlapping if $\Theta \geq \Omega$.
In this case, if we choose
\begin{equation}\label{eq:shannonfilter}
  H(k) =
  \begin{cases}
    \frac{\pi}{\Theta}, & \text{if $|k| < \Omega$} \\
    0, & \text{if $|k| \geq \Omega$},
  \end{cases}
\end{equation}
then $H(k)$ simply projects out everything except the $0^{th}$ copy, and thus \eqref{eq:firstordercopies} is satisfied for any bandlimited $|\phi)$.
This gives the reconstruction kernel
\begin{equation}
  K_n(x) = (x|H|x_n) = \sinc [ \Omega (x-x_n) ].
\end{equation}
In the case where we choose $\Theta = \Omega$, we have $x_n = \frac{n\pi}{\Omega}$, and hence arrive at the Shannon sampling formula \eqref{eq:shannon}.
Note that when we choose $\Theta = \Omega$, the copies $m = \pm 1$ meet the $m = 0$ copy at $k = \pm \Omega$ (respectively).
However, if we constrain our bandlimited functions to vanish at the boundary of the interval, we do not have to be concerned with these points.

If $\Theta$ is chosen strictly larger than $\Omega$, note that $H(k)$ can be chosen arbitrarily in the region $\Omega < |k| < -\Omega + 2\Theta$, since none of the copies of $\tilde{\phi}$ have support there.
Typically this is exploited in practical scenarios so that $H(k)$ can decay to zero more smoothly outside of the bandlimited interval, rather than the sharp drop to zero at $k = \pm \Omega$ in the idealized case.
This can help reduce sources of error which arise in practice \cite{Jerri1977}.
Note, however, that it can only be accomplished by choosing a sample density larger than theoretically necessary.

\begin{figure}[htb]

  \centering
  
  \includegraphics{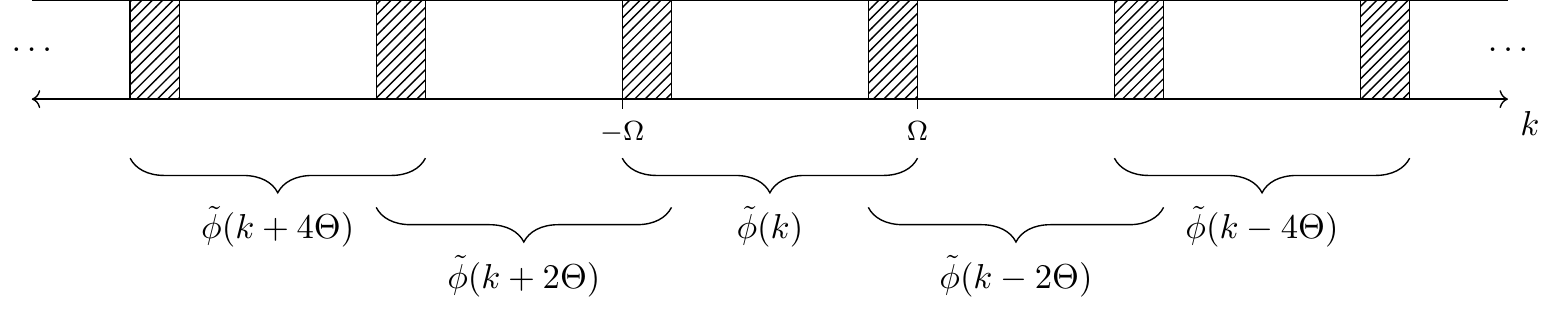}

  \caption{\small An illustration of the support of copies of $\tilde{\phi}(k)$ shifted by integer multiples of $2\Theta$, for $\Theta < \Omega$. In this case, the different copies overlap each other in the regions indicated by the hatching.}
  \label{fig:copies_undersampling}
  
\end{figure}

If we consider choosing $\Theta < \Omega$, as illustrated in Figure~\ref{fig:copies_undersampling}, we will have different copies overlapping each other.
In this situation, it is not possible to choose a $H(k)$ so that \eqref{eq:firstordercopies} is satisfied for every element in the bandlimited subspace.
This is because it is possible to find different bandlimited functions for which the sum in the right-hand side of \eqref{eq:firstordercopies} is the same in the region $|k| < \Omega$ (such functions are called \emph{aliases}).
Therefore, the filter $H(k)$ will not be able to distinguish between these aliases since it simply acts by pointwise multiplication (in $k$-space) outside of the sum in \eqref{eq:firstordercopies}.
We then conclude that we cannot find a reconstruction formula in this way if $\Theta < \Omega$.
This demonstrates the Nyquist density criterion (or that of Beurling-Landau more generally), namely that the minimum density of sample points required for a stable reconstruction occurs when $\Theta = \Omega$.

\section{Temporal sampling on Minkowski spacetime}
\label{sec:minkowski_sampling}

In this section, we will examine the reconstruction properties that one obtains for bandlimitation on Minkowski spacetime, as well as provide a detailed development of an explicit reconstruction formula.
This will be mainly a review of ideas presented in \cite{Kempf2004a} and \cite{KempfChatwindaviesMartin2013}.
We will, however, provide some more details in the development of the reconstruction formula first stated in \cite{KempfChatwindaviesMartin2013}, along with a visualization of how this formula is obtained.
In addition, here we will show that there is some freedom in the reconstruction kernel, as well as a potential instability in the reconstruction, both of which were previously unidentified.
However, the primary motivation for giving a detailed construction here is that it will be useful as background for Section~\ref{sec:sampling_symmetries}, where we will give some new observations regarding the interplay of the reconstruction with the Poincar\'e symmetry of the bandlimited function space.

\subsection{Minkowski bandlimitation}

\begin{figure}[htb]

  \centering
  
  \includegraphics{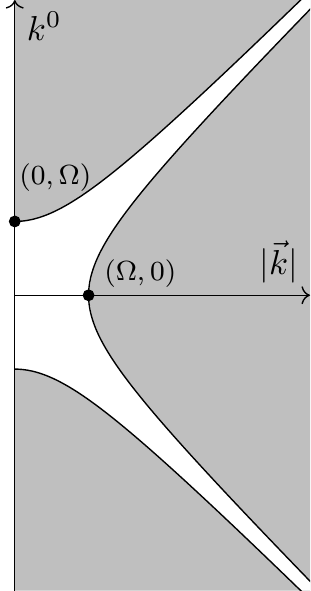}

  \caption{\small $(1+1)$-dimensional representation of the bandlimited region in Fourier space defined by $| (k^0)^2 - \vec{k}^2 | < \Omega^2$.
  Each point in the figure corresponds to a $(N-1)$-sphere associated with the direction of $\vec{k}$.
  The allowed region for the support of the bandlimited functions is indicated in white.
  This region is bounded by hyperbolas with asymptotes $k^0 = \pm |\vec{k}|$.}
  \label{fig:Minkowski_bandlimit}
  
\end{figure}

Recall that the Poincar\'e-invariant notion of bandlimitation on $(N+1)$-dimensional Minkowski spacetime consists of a restriction of the support of fields in Fourier space to the region $|k^2| < \Omega^2$:
\begin{equation}\label{eq:Minkowski_bandlimit}
  \phi(x) = \int_{|k^2| < \Omega^2} \frac{d^{N+1}k}{(2\pi)^{N+1}} \tilde{\phi}(k) e^{-i k \cdot x},
\end{equation}
where $k \cdot x \equiv k^0 x^0 - \vec{k} \cdot \vec{x}$.
The bandlimited region $|k^2| = |(k^0)^2 - \vec{k}^2| < \Omega^2$ is illustrated in Figure~\ref{fig:Minkowski_bandlimit}.

Although the bandlimited region of Fourier space shrinks (eventually to zero) as one travels away from the origin along the lines $k^0 = \pm |\vec{k}|$, the volume of the entire region is infinite (in $1+1$ as well as higher-dimensional Minkowski spacetime).
By our discussion in Subsection~\ref{subsec:sampling_overview}, this would indicate that the Beurling-Landau sample density necessary for the stable reconstruction of bandlimited functions in this space is also infinite.
Note that Landau's result also applies to the Minkowski spacetime case because the metric signature does not play a significant role in the definition of square-integrability nor of Fourier space when the metric is flat.
One could view the extra minus sign in the Fourier phase (for example, in \eqref{eq:Minkowski_bandlimit}) as simply establishing a different Fourier convention for the temporal component.

Therefore, we see that it is not possible to represent this space of functions on a (uniformly) discrete sampling lattice in Minkowski spacetime.
Nevertheless, in \cite{Kempf2004a} it was observed that through a temporal-spatial splitting of the allowed Fourier space region, one can develop certain reconstruction properties for these bandlimited functions.
Consider an arbitrary bandlimited function, $\phi(x)$, and write the \emph{spatial} Fourier transform as $\phi_{\vec{k}}(t)$, i.e., defined so that
\begin{equation}
  \phi(x) = \int \frac{d^N\vec{k}}{(2\pi)^N} \phi_{\vec{k}}(t) e^{i \vec{k} \cdot \vec{x}}.
\end{equation}
Now if we examine a fixed spatial mode $\vec{k}$, in order for the function $\phi(x)$ to be bandlimited in the Poincar\'e-invariant sense, the time-varying amplitude $\phi_{\vec{k}}(t)$ must be bandlimited in time.
Specifically, the temporal frequencies, $k^0$, of the amplitude $\phi_{\vec{k}}(t)$ of fixed $\vec{k}$, must satisfy:
\begin{equation}
  | (k^0)^2 - \vec{k}^2 | < \Omega^2 \iff \max \lbrace 0, \vec{k}^2 - \Omega^2 \rbrace < (k^0)^2 < \vec{k}^2 + \Omega^2.
\end{equation}
If we define $r_\pm(\vec{k}) := \sqrt{ \vec{k}^2 \pm \Omega^2 }$, then we can write the set of allowed temporal frequencies for a fixed $\vec{k}$ as:
\begin{equation}
  k^0 \in I(\vec{k}) :=
  \begin{cases}
    ( -r_+(\vec{k}), r_+(\vec{k}) ), & \text{if $|\vec{k}| \leq \Omega$} \\
    \!\begin{aligned}
      & ( -r_+(\vec{k}), -r_-(\vec{k}) ) \cup ( r_-(\vec{k}), r_+(\vec{k}) ),
    \end{aligned}
    & \text{if $|\vec{k}| > \Omega$}.
  \end{cases}
\end{equation}
The intervals for different choices of $\vec{k}$ can be visualized as vertical sections in Figure~\ref{fig:Minkowski_bandlimit}.

For any fixed $\vec{k}$, there is a finite temporal bandwidth (size of the interval $I(\vec{k})$) for the time-dependence of the Fourier coefficient $\phi_{\vec{k}}(t)$.
Therefore, one expects to be able to reconstruct $\phi_{\vec{k}}(t)$ from sample values at a discrete set of points in time, provided that the sample density is greater than the Beurling-Landau density corresponding to the mode $\vec{k}$.
Note that this does not contradict the requirement of an infinite sample density required for the reconstruction of the full function $\phi(x)$, since we are only considering a fixed mode $\vec{k}$.
For a spatial mode with $|\vec{k}| \leq \Omega$, $\phi_{\vec{k}}(t)$ is bandlimited to $k^0 \in ( -r_+(\vec{k}), r_+(\vec{k}) )$, which can be identified with the one-dimensional Shannon case.
Therefore, for this mode we can use \eqref{eq:shannon} as the reconstruction formula, where we identify the bandlimit as $r_+(\vec{k})$ and reconstruct $\phi_{\vec{k}}(t)$ from samples at the points $\{ t_{\vec{k},n} = n\pi / r_+(\vec{k}) \}_{n \in \mathbb{Z}}$.
For spatial modes with $|\vec{k}| > \Omega$, the function $\phi_{\vec{k}}(t)$ is bandlimited to $k^0 \in ( -r_+(\vec{k}), -r_-(\vec{k}) ) \cup ( r_-(\vec{k}), r_+(\vec{k}) )$.
An explicit reconstruction formula is also known for this case of the union of two intervals using samples at the Beurling-Landau density $[ r_+(\vec{k}) - r_-(\vec{k}) ] / \pi$ \cite{Kohlenberg1953}.
We will employ the method of \cite{Kohlenberg1953} in the next subsection, and simply refer the reader to this reference for the one-dimensional case which we will not need to state here.
The point is that in both cases $|\vec{k}| \leq \Omega$ and $|\vec{k}| > \Omega$, there is a known reconstruction formula for $\phi_{\vec{k}}(t)$ from samples in time taken at the minimal density.

How does the sample density change with $\vec{k}$?
For the spatial mode $\vec{k} = 0$, we have a temporal bandwidth of $2\Omega$.
As we increase $|\vec{k}|$, the temporal bandwidth increases to a maximum of $2\sqrt{2}\Omega$ for spatial modes with $|\vec{k}| = \Omega$.
The bandwidth then decreases monotonically, and tends to zero as $|\vec{k}| \to \infty$.
Therefore, fixed spatial modes with $|\vec{k}| = \Omega$ (i.e., at the Planck scale) require the largest temporal sample density.
Modes with much larger spatial frequencies require a temporal sampling which is much less dense, and which decreases to zero density as $|\vec{k}| \to \infty$.
Thus, although these bandlimited functions can have arbitrarily small wavelengths, those shorter than the Planck length have a small temporal bandwidth.

Now that we know there is a reconstruction formula for each fixed spatial mode $\phi_{\vec{k}}(t)$, can we combine them into a reconstruction formula for $\phi(x)$?
Schematically, let us write the reconstruction of mode $\phi_{\vec{k}}(t)$ from an appropriate lattice $\{ t_{\vec{k},n} \}_{n \in \mathbb{Z}}$ as
\begin{equation}
  \phi_{\vec{k}}(t) = \sum_{n \in \mathbb{Z}} K_{\vec{k},n}(t) \phi_{\vec{k}}(t_{\vec{k},n}),
\end{equation}
where $K_{\vec{k},n}(t)$ is an appropriate reconstruction kernel (which is generally different for each $|\vec{k}|$).
Then we can write the full function as
\begin{equation}
\label{eq:Minkowksi_reconstruction_permode}
  \phi(x) = \int \frac{d^N\vec{k}}{(2\pi)^N} \sum_{n \in \mathbb{Z}} K_{\vec{k},n}(t) \phi_{\vec{k}}(t_{\vec{k},n}) e^{i \vec{k} \cdot \vec{x}}.
\end{equation}
In order to turn this into a sampling of $\phi(x)$ at a discrete set of times, we would like to do an inverse spatial Fourier transform of $\phi_{\vec{k}}(t_{\vec{k},n})$.
However, the mode-dependence of the sample times $t_{\vec{k},n}$ is an obstruction to identifying $\phi_{\vec{k}}(t_{\vec{k},n})$ as the spatial Fourier transform of $\phi(x)$ at some time.
To overcome this, we notice that because there is a maximum temporal bandwidth of $2\sqrt{2}\Omega$ (for modes with $|\vec{k}| = \Omega$), we could try to eliminate the mode dependence of the sample times by sampling all of the modes at the density corresponding to this bandwidth.
The expense is that, for most of the modes, this sampling is inefficient as they are being sampled at a density larger than required.
However, the gain is that if we can find a reconstruction of all of the $\phi_{\vec{k}}(t)$'s from some single temporal lattice $\{ t_n \}_{n \in \mathbb{Z}}$, then this would allow us to write:
\begin{align}
  \phi(x) &= \int \frac{d^N\vec{k}}{(2\pi)^N} \sum_{n \in \mathbb{Z}} K_{\vec{k},n}(t) \phi_{\vec{k}}(t_n) e^{i \vec{k} \cdot \vec{x}} \\
  &= \sum_{n \in \mathbb{Z}} \int d^N\vec{x}' \left[ \int \frac{d^N\vec{k}}{(2\pi)^N} K_{\vec{k},n}(t) e^{i \vec{k} \cdot ( \vec{x} - \vec{x}' ) } \right] \phi(t_n,\vec{x}'). \label{eq:Minkowksi_reconstruction_allmode}
\end{align}
The function in the square brackets is a kind of reconstruction kernel, depending on $n$, $t$, and $\vec{x}-\vec{x}'$.
This allows for a reconstruction of the bandlimited function $\phi(x)$ from samples consisting of the functions $\phi(t_n,\vec{x})$ on a discrete set of constant-time hypersurfaces.
Although this is not a sampling and reconstruction of the traditional kind as in \eqref{eq:shannon} and \eqref{eq:reconstruction}, we expect to have a kind of reconstruction from samples in time, by sampling all of the spatial modes at the same density of $\sqrt{2}\Omega/\pi$.
Note that despite being able to sample in time, the spacetime sample density remains infinite because the samples require knowledge of the function on the entire constant-time hypersurfaces, and these functions can have arbitrarily short wavelengths.

What about an analogous sampling in space (rather than in time)?
It turns out there is an analogue of \eqref{eq:Minkowksi_reconstruction_permode}, but not of \eqref{eq:Minkowksi_reconstruction_allmode}, for a sampling in space in dimensions $3+1$ and higher.
Suppose for an arbitrary bandlimited function, $\phi(x)$, we write the \emph{temporal} Fourier transform as $\phi_{k^0}(\vec{x})$, so that
\begin{equation}
  \phi(x) = \int \frac{dk^0}{2\pi} \phi_{k^0}(\vec{x}) e^{-i k^0 t}.
\end{equation}
Now we examine a fixed temporal mode $k^0$.
This mode has spatial dependence $\phi_{k^0}(\vec{x})$, and, similar to above, in order for $\phi(x)$ to be bandlimited, the spatial frequencies of $\phi_{k^0}(\vec{x})$ must satisfy:
\begin{equation}
  |\vec{k}| \in I(k^0) :=
  \begin{cases}
    [ 0, r_+(k^0) ), & \text{if $|k^0| \leq \Omega$} \\
    ( r_-(k^0), r_+(k^0) ), & \text{if $|k^0| > \Omega$}.
  \end{cases}
\end{equation}
The intervals for different choices of $k^0$ can be visualized as horizontal sections in Figure~\ref{fig:Minkowski_bandlimit}.

In order to determine the spatial sample density required to reconstruct a fixed temporal mode, $\phi_{k^0}(\vec{x})$, the Beurling-Landau condition tells us to find the volume of its support in $\vec{k}$-space.
Unlike the case of fixing a spatial mode, the volume of the $\vec{k}$-space region $|\vec{k}| \in I(k^0)$ depends on the spatial dimension $N$, since we are only restricting the magnitude of $\vec{k}$ to lie in $I(k^0)$.
The volume for each fixed $k^0$ is:
\begin{equation}
  \text{Vol}(I(k^0)) =
  \begin{cases}
    V_N r_+(k^0)^N, & \text{if $|k^0| \leq \Omega$} \\
    V_N ( r_+(k^0)^N - r_-(k^0)^N ), & \text{if $|k^0| > \Omega$},
  \end{cases}
\end{equation}
where $V_N = \pi^{N/2}/\Gamma(N/2+1)$ is the volume of the unit ball in $N$ dimensions.
We see that for every fixed $k^0$, this volume is finite, and thus we expect to be able to reconstruct the temporal mode $\phi_{k^0}(\vec{x})$ from a discrete spatial lattice with density proportional to $\text{Vol}(I(k^0))$.
Since this can be done for each of the temporal modes, we get an analogue of \eqref{eq:Minkowksi_reconstruction_permode} for sampling in space.

In order to get a reconstruction formula analogous to \eqref{eq:Minkowksi_reconstruction_allmode}, we need to be able to sample all of the modes from the same lattice in space.
For this to be possible, we need to find a spatial sample density which is large enough to reconstruct all of the temporal modes.
In $1+1$ dimensions, as we saw for the spatial modes, $\text{Vol}(I(k^0))$ has an upper bound of $2 \sqrt{2} \Omega$ which occurs at $|k^0| = \Omega$.
In $2+1$ dimensions, $\text{Vol}(I(k^0))$ has a maximum of $2 \pi \Omega^2$, which occurs for all $|k^0| \geq \Omega$.
However, in $3+1$ dimensions and higher, $\text{Vol}(I(k^0))$ diverges as $|k^0| \to \infty$, thus the required spatial sample density increases without bound for modes of large temporal frequency.
Therefore, we see that although we should be able to reconstruct each temporal mode from a discrete spatial lattice, there is no uniformly discrete spatial lattice of sufficiently high density to reconstruct all of the temporal modes in $3+1$ dimensions and higher, and hence no analogue of \eqref{eq:Minkowksi_reconstruction_allmode} in these cases.

For the case of temporal sampling, we have so far only argued that we expect there to be a reconstruction formula of the form \eqref{eq:Minkowksi_reconstruction_allmode}, but we have not determined explicitly how all of the spatial modes can be sampled from the same temporal lattice.
In Subsection~\ref{subsec:Minkowski_reconstruction}, we will develop an explicit reconstruction formula for \eqref{eq:Minkowksi_reconstruction_allmode}, demonstrating that it is possible to reconstruct the full function $\phi(x)$ from a discrete set of constant-time hypersurfaces.
We will see that some care is required in choosing a sampling lattice from which all of the modes can be reconstructed simultaneously.
We will focus on the case of temporal sampling since a spatial sampling of $\phi(x)$ of the form \eqref{eq:Minkowksi_reconstruction_allmode} is not possible in $3+1$ dimensions.

\subsection{Reconstruction formula}
\label{subsec:Minkowski_reconstruction}

In this subsection, we will produce a temporal lattice and a reconstruction formula which will show that an arbitrary Minkowski-bandlimited function, $\phi(x)$, can be reconstructed from samples $\phi(t_n,\vec{x})$, hence explicitly demonstrating \eqref{eq:Minkowksi_reconstruction_allmode}.

\begin{figure}[htb]

  \centering
  
  \includegraphics{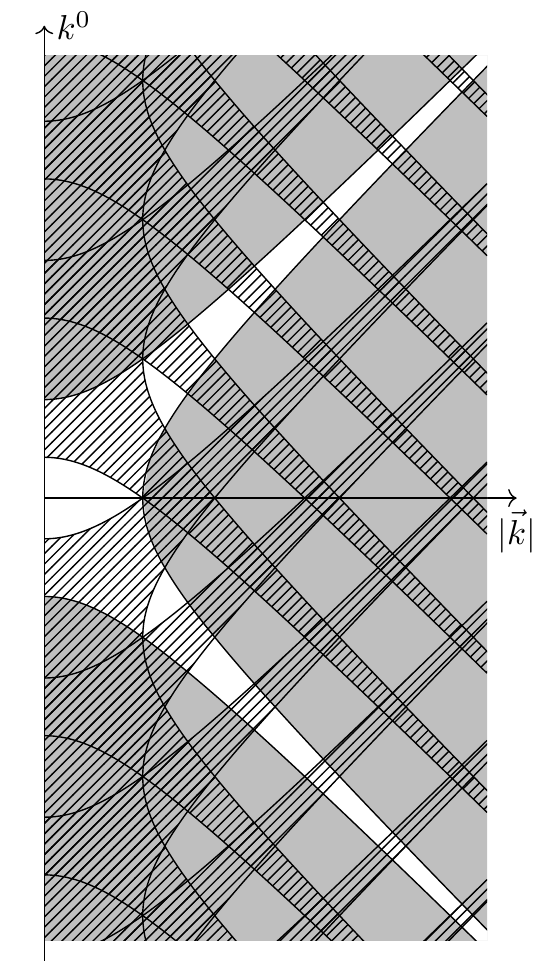}

  \caption{\small A $(1+1)$-dimensional representation of the support of copies of $\tilde{\phi}(k)$ shifted by integer multiples of $2\Theta$ in the $k^0$-direction. The original ($0^{th}$) copy is indicated in white, and all of the other copies are depicted by the hatched regions. For this figure, we chose $\Theta = \Omega/\sqrt{2}$.}
  \label{fig:Minkowski_copies}
  
\end{figure}

First, let us argue that it is not possible to employ a lattice of the form $\{ t_n = n\pi/\Theta \}_{n \in \mathbb{Z}}$ (for any $\Theta \in \mathbb{R}^+$) using a similar construction to Subsection~\ref{subsec:shannon_formula}.
Suppose we were to choose a lattice of this form.
The corresponding sampling operator for this lattice is:
\begin{equation}
  S = \sum_{n \in \mathbb{Z}} |t_n)(t_n| \otimes 1 = \sum_{n \in \mathbb{Z}} \int d^N\vec{x} |t_n,\vec{x})(t_n,\vec{x}| = \sum_{n \in \mathbb{Z}} \int \frac{d^N\vec{k}}{(2\pi)^N} |t_n,\vec{k})(t_n,\vec{k}|.
\end{equation}
We will also define a filter:
\begin{equation}
  H = \int \frac{d^{N+1}k}{(2\pi)^{N+1}} H(k) |k)(k|.
\end{equation}
Recall from Subsection~\ref{subsec:shannon_formula} that the aim is to find a $\Theta$ and a function $H(k)$ so that $HS$ is the identity on the bandlimited subspace, i.e., $|\phi) = HS|\phi)$ for all bandlimited $|\phi)$.
In Fourier space, this can be written
\begin{align}
  \tilde{\phi}(k) \equiv (k|\phi) &\stackrel{!}{=} (k| HS |\phi) \\
  &= H(k) (k| \left[ \sum_{n \in \mathbb{Z}} \int \frac{d^N\vec{k}''}{(2\pi)^N} |t_n,\vec{k}'')(t_n,\vec{k}''| \right] \left[ \int \frac{d^{N+1}k'}{(2\pi)^{N+1}} |k')(k'| \right] |\phi) \\
  &= H(k) \sum_{n \in \mathbb{Z}} \int \frac{d^N\vec{k}''}{(2\pi)^N} \int \frac{d^{N+1}k'}{(2\pi)^{N+1}} e^{i k_0 t_n} (2\pi)^N \delta(\vec{k}-\vec{k}'') e^{-i k'_0 t_n} (2\pi)^N \delta(\vec{k}''-\vec{k}') \tilde{\phi}(k') \\
  &= H(k) \int \frac{dk'_0}{2\pi} \left[ \sum_{n \in \mathbb{Z}} e^{i (k_0 - k'_0) t_n} \right] \tilde{\phi}(k'_0,\vec{k}) \\
  &= H(k) \int \frac{dk'_0}{2\pi} \left[ 2\Theta \sum_{m \in \mathbb{Z}} \delta( k_0 - k'_0 - 2\Theta m ) \right] \tilde{\phi}(k'_0, \vec{k}) \\
  &= \frac{\Theta}{\pi} H(k) \sum_{m \in \mathbb{Z}} \tilde{\phi}(k_0 - 2\Theta m, \vec{k}).
\end{align}
Note our Fourier convention is $(x|k) = e^{-i k \cdot x} = e^{-i k^0 t + i \vec{k} \cdot \vec{x}}$, and we have again used
\begin{equation}
  \sum_{n \in \mathbb{Z}} e^{i (k_0 - k'_0) \frac{n\pi}{\Theta}} = 2\Theta \sum_{m \in \mathbb{Z}} \delta( k_0 - k'_0 - 2\Theta m )
\end{equation}
since we have assumed $t_n = n\pi/\Theta$.
Then, similarly to the one-dimensional case, we have the condition
\begin{equation}
  \tilde{\phi}(k_0,\vec{k}) \stackrel{!}{=} \frac{\Theta}{\pi} H(k) \sum_{m \in \mathbb{Z}} \tilde{\phi}(k_0 - 2\Theta m, \vec{k}).
\end{equation}
Figure~\ref{fig:Minkowski_copies} provides an illustration of the sum on the right-hand side of this equation.

In Subsection~\ref{subsec:shannon_formula}, we argued that in order for the filter to be able to recover the original bandlimited function from the result of the sum over copies, the support of the copies $m \neq 0$ cannot overlap the support of the original ($0^{th}$) copy.
In the one-dimensional case this could be achieved by requiring $\Theta \geq \Omega$.
However, in the present case, it should be clear from Figure~\ref{fig:Minkowski_copies} (and can also be shown analytically) that each of the copies will have some overlap with the original, regardless of how they are translated.
Therefore, we cannot obtain a reconstruction formula using the above procedure in the form presented.

Fortunately, an extension of the above method was devised by Kohlenberg \cite{Kohlenberg1953} which applies when the number of copies which overlap in any given region is bounded. 
In \cite{Kohlenberg1953}, the aim was to develop a reconstruction formula for one-dimensional functions bandlimited to a region in Fourier space of the form $0 < \Omega_1 < |k| < \Omega_2$ (called \emph{bandpass} functions), for which overlap between copies generically occurs when sampling at the minimal density $(\Omega_2 - \Omega_1)/\pi$.
The idea of \cite{Kohlenberg1953} is to consider multiple sampling lattices, $\{ x_n^{(i)} \}_{n \in \mathbb{Z}}$ (where the different lattices are indexed by $i$), with corresponding sampling operators, $S_i$, as well as multiple filters, $H_i$.
A $p^{th}$-order sampling is then said to be possible if one can design a set of lattices and filters so that $\sum_{i=1}^p H_i S_i$ is the identity on the bandlimited function space.

We will show that a temporal sampling and reconstruction of a Minkowski-bandlimited function, $\phi(x)$, can be achieved using a second-order sampling.
Recall that the maximum temporal bandwidth of the spatial modes of $\phi(x)$ was found to be $2\sqrt{2}\Omega$, and thus the Beurling density we expect to require for the temporal sampling lattice is $\sqrt{2}\Omega/\pi$.
Instead of a single equidistantly-spaced lattice with this density, we will consider two equidistantly-spaced lattices each with half the density, $\Omega/\pi \sqrt{2}$ (so that the union has a Beurling density of $\sqrt{2}\Omega/\pi$), and which are shifted from each other by an amount $\tau$ (which is unspecified at this stage):
\begin{equation}
  t_n^{(1)} = \frac{n \pi}{\Theta}, \quad t_n^{(2)} = \frac{n \pi}{\Theta} + \tau, \qquad (n \in \mathbb{Z}),
\end{equation}
where we fix $\Theta = \Omega/\sqrt{2}$.
Apart from the abstract expectation, it will become clear during the following analysis that this is the minimal density which can be chosen in order to obtain a reconstruction formula using this method.
We have sampling operators corresponding to the two lattices,
\begin{equation}
  S_i := \sum_{n \in \mathbb{Z}} \int \frac{d^N\vec{k}}{(2\pi)^N} |t_n^{(i)},\vec{k})(t_n^{(i)},\vec{k}|,
\end{equation}
as well as two filters,
\begin{equation}
  H_i := \int \frac{d^{N+1}k}{(2\pi)^{N+1}} H_i(k) |k)(k|.
\end{equation}

Now, similar to before, the aim is to construct $H_1(k)$ and $H_2(k)$ so that for any bandlimited $\phi$, we have:
\begin{equation}
  |\phi) \stackrel{!}{=} (H_1 S_1 + H_2 S_2) |\phi)
\end{equation}
(recall we have already fixed $\Theta = \Omega/\sqrt{2}$, but still have freedom in our choice of $\tau$).
Using the same steps as before, we can write this in Fourier space as:
\begin{align}
  \tilde{\phi}(k) &= (k|\phi) \\
  &\stackrel{!}{=} (k| ( H_1 S_1 + H_2 S_2 ) |\phi) \\
  &= \frac{\Theta}{\pi} H_1(k) \sum_{m \in \mathbb{Z}} \tilde{\phi}(k_0 - 2\Theta m, \vec{k}) + \frac{\Theta}{\pi} H_2(k) \sum_{m \in \mathbb{Z}} e^{ i 2 \Theta \tau m } \tilde{\phi}(k_0 - 2\Theta m, \vec{k}) \\
  &= \frac{\Theta}{\pi} \sum_{m \in \mathbb{Z}} \left[ H_1(k) + e^{ i 2 \Theta \tau m } H_2(k) \right] \tilde{\phi}(k_0 - 2\Theta m, \vec{k}), \label{eq:minkowskisecondordercopies}
\end{align}
where we used
\begin{equation}
  \sum_{n \in \mathbb{Z}} e^{i (k_0 - k'_0) \left( \frac{n\pi}{\Theta} + \tau \right)} = 2\Theta \sum_{m \in \mathbb{Z}} e^{ i 2 \Theta \tau m } \delta( k_0 - k'_0 - 2\Theta m ).
\end{equation}
The goal is to design the filters to isolate the $m=0$ term in \eqref{eq:minkowskisecondordercopies} and set the remaining terms to zero.
This would be simple if the filters depended on the index $m$, however, they only depend on $k^0$ and $\vec{k}$.

The design strategy will be to first examine fixed regions of Fourier space and determine which copies have support there.
Let us begin by fixing an arbitrary $\vec{k}$ with $|\vec{k}| \leq \Omega$.
Then \eqref{eq:minkowskisecondordercopies} simply reduces to a one-dimensional problem in $k^0$.
For the chosen $\vec{k}$, $\tilde{\phi}(k^0,\vec{k})$ is bandlimited to $k^0 \in (-r_+(\vec{k}),r_+(\vec{k}))$, and the $m^{th}$ copy is bandlimited to $k^0 \in (-r_+(\vec{k}) + 2\Theta m, r_+(\vec{k}) + 2\Theta m)$.
For our choice of $\Theta = \Omega/\sqrt{2}$, the $k^0$-intervals for different $m$'s will be overlapping.
This is similar to the situation depicted in Figure~\ref{fig:copies_undersampling}, which can be directly compared with a vertical section of Figure~\ref{fig:Minkowski_copies} at some fixed $|\vec{k}| \leq \Omega$.
Since we are only interested in isolating the $0^{th}$ copy, we can focus on the region $k^0 \in (-r_+(\vec{k}),r_+(\vec{k}))$ and set $H_1(k^0,\vec{k}) = H_2(k^0,\vec{k}) = 0$ outside of this interval (for this particular $\vec{k}$).
Which copies have support in $k^0 \in (-r_+(\vec{k}),r_+(\vec{k}))$?
Copy $m$ will have support in this region if either its left endpoint lies in the interval, $-r_+(\vec{k}) \leq -r_+(\vec{k}) + 2\Theta m < r_+(\vec{k})$, or the right endpoint, $-r_+(\vec{k}) < r_+(\vec{k}) + 2\Theta m \leq r_+(\vec{k})$.
For $\Theta = \Omega/\sqrt{2}$ (and $|\vec{k}| \leq \Omega$), this occurs only for $m = -1, 0, \text{ and } 1$.
Let us focus on the interval $k^0 \in (-r_+(\vec{k}) + 2\Theta, r_+(\vec{k}))$ where copies $0$ and $1$ overlap (once can check that copies $1$ and $-1$ do not have overlapping support).
Then for our fixed $\vec{k}$, and $k^0$ in this interval, the condition \eqref{eq:minkowskisecondordercopies} becomes
\begin{equation}
  \tilde{\phi}(k) \stackrel{!}{=} \frac{\Theta}{\pi} \left[ H_1(k) + H_2(k) \right] \tilde{\phi}(k_0,\vec{k}) + \frac{\Theta}{\pi} \left[ H_1(k) + e^{i 2 \Theta \tau} H_2(k) \right] \tilde{\phi}(k_0 - 2\Theta, \vec{k}).
\end{equation}
Clearly this condition can be met if the filters satisfy
\begin{align}
  &\frac{\Theta}{\pi} \left[ H_1(k) + H_2(k) \right] \stackrel{!}{=} 1, \\
  &\frac{\Theta}{\pi} \left[ H_1(k) + e^{i 2 \Theta \tau} H_2(k) \right] \stackrel{!}{=} 0,
\end{align}
which are solved by
\begin{align}
  &H_1(k) = \frac{\pi}{\Theta} (1 - e^{-i 2 \Theta \tau})^{-1}, \\
  &H_2(k) = \frac{\pi}{\Theta} (1 - e^{i 2 \Theta \tau})^{-1},
\end{align}
provided $\tau \not\in \tfrac{\pi}{\Theta} \mathbb{Z}$.
We see that despite the fact that different copies overlap each other, because we have two filters, we can isolate the original copy in the overlapping region.
Similarly, the $0$ and $-1$ copies overlap in the interval $k^0 \in (-r_+(\vec{k}), r_+(\vec{k}) - 2\Theta)$, and one can check that we can satisfy our equation \eqref{eq:minkowskisecondordercopies} in this region with
\begin{align}
  &H_1(k) = \frac{\pi}{\Theta} (1 - e^{i 2 \Theta \tau})^{-1}, \\
  &H_2(k) = \frac{\pi}{\Theta} (1 - e^{-i 2 \Theta \tau})^{-1}.
\end{align}
For the remainder of the interval, i.e., $k^0 \in (r_+(\vec{k}) - 2\Theta, -r_+(\vec{k}) + 2\Theta)$ (which is nonempty, except when $|\vec{k}| = \Omega$ for which two endpoints meet), only copy $0$ is present and so we only have the $m=0$ term in the sum.
Therefore, in this interval we only get the constraint
\begin{equation}
  \frac{\Theta}{\pi} \left[ H_1(k) + H_2(k) \right] \stackrel{!}{=} 1.
\end{equation}
Hence, for each fixed $k$ in this region, we have a one-dimensional space of solutions for the filters.
We could choose some arbitrary solution, but we could also simply impose either of the additional constraints that we imposed when the $m = 0$ and $m = \pm 1$ terms were overlapping.
For symmetry, let us split the interval at the origin, and impose the additional constraint $\frac{\Theta}{\pi} \left[ H_1(k) + e^{i 2 \Theta \tau} H_2(k) \right] \stackrel{!}{=} 0$ on the right, and $\frac{\Theta}{\pi} \left[ H_1(k) + e^{-i 2 \Theta \tau} H_2(k) \right] \stackrel{!}{=} 0$ on the left.
Combining everything for the case of $|\vec{k}| \leq \Omega$, we have shown that by choosing
\begin{equation}\label{eq:filtersoln1}
  H_1(k^0,\vec{k}) = H_2(-k^0,\vec{k}) = 
  \begin{cases}
    \frac{\pi}{\Theta} (1 - e^{-i 2 \Theta \tau})^{-1}, & \text{if $k^0 \in [0, r_+(\vec{k}))$} \\
    \frac{\pi}{\Theta} (1 - e^{i 2 \Theta \tau})^{-1}, & \text{if $k^0 \in (-r_+(\vec{k}), 0)$} \\
    0, & \text{otherwise},
  \end{cases}
\end{equation}
we have $\tilde{\phi}(k) = (k|(H_1 S_1 + H_2 S_2)|\phi)$ in the region $|\vec{k}| \leq \Omega$ for any bandlimited $\phi$.
Notice that, as we have indicated in this expression, the solutions we have found above satisfy $H_1(k^0,\vec{k}) = H_2(-k^0,\vec{k})$.
We also note that these functions are piecewise constant.

It is straightforward to show that if we had chosen the lattices to be less dense, i.e., $\Theta < \Omega/\sqrt{2}$, then there would be a $|\vec{k}| \leq \Omega$ for which we would have the copies $m = 0, 1, \text{ and } 2$ all overlapping in the interval $k^0 \in (-r_+(\vec{k}), r_+(\vec{k}))$.
In the region where they all overlap, \eqref{eq:minkowskisecondordercopies} yields three constraints for the two filters which are not simultaneously satisfiable.
If one considered third-order sampling, a third filter could be used along with the other two to satisfy these three constraints.
However, this third filter would come along with a third lattice, which would increase the overall sample density.
Therefore, including additional filters by increasing the order of the sampling can help to disentangle multiple overlaps, but will not necessarily aid in decreasing the total sample density.
This supports the expectation that the lattices we have chosen are at the minimal density needed for reconstruction.

\begin{figure}[htb]

  \centering
  
  \includegraphics{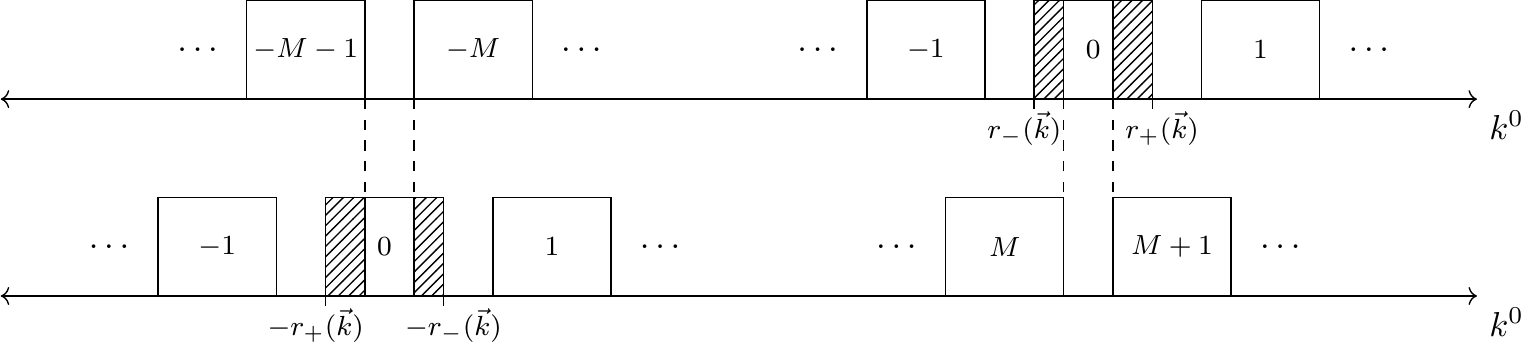}
  
  \caption{\small An illustration of the support of copies of $\tilde{\phi}(k^0,\vec{k})$ for some fixed $|\vec{k}| > \Omega$.
  The copies are shifted by integer multiples of $2\Theta = \sqrt{2}\Omega$.
  The copies of the positive band are illustrated in the upper part of the figure, and the copies of the negative band in the lower part.
  The hatching shows the regions where there is overlap between the $0^{th}$ copy and copies $m = \pm M, \pm (M+1)$.}
  \label{fig:Minkowski_copies_bandpass}
  
\end{figure}

Now let us proceed to examine the remaining cases of fixed $\vec{k}$ with $|\vec{k}| > \Omega$.
This is done using the same method, only that determining where the overlaps occur between copies (and hence which terms contribute in the sum in \eqref{eq:minkowskisecondordercopies}) is somewhat more involved.
We will present it here for completeness.
First, recall that for a fixed $|\vec{k}| > \Omega$, a bandlimited $\tilde{\phi}(k^0,\vec{k})$ has support in $k^0 \in (-r_+(\vec{k}), -r_-(\vec{k})) \cup (r_-(\vec{k}), r_+(\vec{k}))$.
Let us call the interval $(r_-(\vec{k}), r_+(\vec{k}))$ the \emph{positive band}, and $(-r_+(\vec{k}), -r_-(\vec{k}))$ the \emph{negative band} of temporal frequencies for a fixed $\vec{k}$.
Now, each copy, $\tilde{\phi}(k^0 - 2\Theta m, \vec{k})$, has both a positive and negative band shifted by $2 \Theta m$ in the $k^0$-direction.
Provided that $\Theta > \Omega/\sqrt{2}$, none of the copies of the positive band overlap one another, and none of the copies of the negative band overlap one another.
However, it is possible that one or more copies of the negative band will overlap a given copy of the positive band, and vice versa.
This is illustrated in Figure~\ref{fig:Minkowski_copies_bandpass}, which can be viewed as a vertical section of Figure~\ref{fig:Minkowski_copies} for some particular $|\vec{k}| > \Omega$.

We need to determine in which regions we have overlap between copy $0$ and other copies, as well as find solutions for the filters in these regions which can isolate the $m=0$ term in the sum of \eqref{eq:minkowskisecondordercopies}.
We will tackle the second problem first.
Consider some point $k = (k^0,\vec{k})$ where the $m=0$ and $m=M$ copies overlap, the condition \eqref{eq:minkowskisecondordercopies} becomes
\begin{align}
  \tilde{\phi}(k) \stackrel{!}{=} \frac{\Theta}{\pi} [ H_1(k) + H_2(k) ] \tilde{\phi}(k_0,\vec{k}) + \frac{\Theta}{\pi} [ H_1(k) + e^{i 2 \Theta \tau M} H_2(k) ] \tilde{\phi}(k_0 - 2\Theta M, \vec{k}).
\end{align}
Similar to before, this is uniquely solved by choosing
\begin{align}
  &H_1(k) = \frac{\pi}{\Theta} ( 1 - e^{-i 2 \Theta \tau M} )^{-1}, \label{eq:filtersolnoverlapM_1} \\
  &H_2(k) = \frac{\pi}{\Theta} ( 1 - e^{i 2 \Theta \tau M} )^{-1}, \label{eq:filtersolnoverlapM_2}
\end{align}
as long as $\tau \not\in \frac{\pi}{M \Theta} \mathbb{Z}$.
If three (or more) copies overlap at some point, then we will again obtain three (or more) constraints for the filters which cannot all be satisfied.\footnote{
If we have overlap between copies $0$, $M_1$, and $M_2$, then it turns out the three constraints can all be satisfied if we choose $\tau \in \frac{\pi}{(M_2-M_1) \Theta} \mathbb{Z}$.
However, we will see below that we must choose $\tau \in \frac{\pi}{\Theta} ( \mathbb{R} \setminus \mathbb{Q} )$, since for every $M \in \mathbb{Z} \setminus \{0\}$ there will always be a region of Fourier space where we need the solutions \eqref{eq:filtersolnoverlapM_1} and \eqref{eq:filtersolnoverlapM_2}, which requires $\tau \not\in \frac{\pi}{M \Theta} \mathbb{Z}$.
}
Since the condition $\Theta > \Omega/\sqrt{2}$ guarantees that none of the copies of the positive band overlap one another, and none of the copies of the negative band overlap one another, we cannot have more than two copies overlapping at any given point.
Therefore, the situation where three or more copies are overlapping does not occur, and we can conclude that employing a second-order sampling should be sufficient.
There will, however, be regions where only the $m=0$ copy is supported.
At each of these points, we again have a one-dimensional space of solutions for the filters.
Below we will choose to employ the solutions \eqref{eq:filtersolnoverlapM_1} and \eqref{eq:filtersolnoverlapM_2} for some particular choices of $M$ at these points, in a way that simplifies the final expressions.

Now that we know how the filters should be chosen at points where there are copies which overlap the $m=0$ copy, we simply need to determine where these overlaps occur.
The $m^{th}$ copy of the negative band will overlap the $0^{th}$ copy of the positive band if either its left endpoint lies within the positive band, $r_-(\vec{k}) \leq -r_+(\vec{k}) + 2\Theta m < r_+(\vec{k})$, or its right endpoint, $r_-(\vec{k}) < -r_-(\vec{k}) + 2\Theta m \leq r_+(\vec{k})$.
At least one of these conditions will be met if $\vec{k}$ is in the range $-\Omega^2 + \Theta^2 m^2 < \vec{k}^2 < \Omega^2 + \Theta^2 m^2$ for $m \geq 0$.
For example, the $m=0$ and $m=1$ copies overlap if $\vec{k}^2 < \tfrac32 \Omega^2$, and the $m=0$ and $m=2$ copies overlap if $\Omega^2 < \vec{k}^2 < 3 \Omega^2$.
Therefore, if $\Omega < |\vec{k}| < \sqrt{3/2} \Omega$, we have overlap between the $m=0$ and $m=1$ copies, as well as the $m=0$ and $m=2$ copies.
The first overlap occurs for $k^0 \in (r_-(\vec{k}), -r_-(\vec{k}) + 2\Theta)$, and the second for $k^0 \in (-r_+(\vec{k}) + 4\Theta, r_+(\vec{k}))$.
These are disjoint because, as we said above, different copies of the negative band do not overlap one another.
Thus, for each $\vec{k}$ in the range $\Omega < |\vec{k}| < \sqrt{3/2} \Omega$, we choose the filters to be \eqref{eq:filtersolnoverlapM_1} and \eqref{eq:filtersolnoverlapM_2} with $M=1$ for $k^0 \in (r_-(\vec{k}), -r_-(\vec{k}) + 2\Theta)$, and \eqref{eq:filtersolnoverlapM_1} and \eqref{eq:filtersolnoverlapM_2} with $M=2$ for $k^0 \in (-r_+(\vec{k}) + 4\Theta, r_+(\vec{k}))$.
For $k^0 \in (-r_-(\vec{k}) + 2\Theta, -r_+(\vec{k}) + 4\Theta)$, which is nonempty for $\Omega < |\vec{k}| < \sqrt{3/2} \Omega$, we have only the $m=0$ copy.
We will choose the filters as \eqref{eq:filtersolnoverlapM_1} and \eqref{eq:filtersolnoverlapM_2} with $M=2$ here.

Before proceeding to analyze copies $m > 2$, let us first note that the overlap between the $m^{th}$ copy of the positive band and the $0^{th}$ copy of the negative band is simply a mirrored version of the situation just considered.
Here, the condition for an overlap is $-\Omega^2 + \Theta^2 m^2 < \vec{k}^2 < \Omega^2 + \Theta^2 m^2$ and $m \leq 0$.
Therefore, in the same range $\Omega < |\vec{k}| < \sqrt{3/2} \Omega$ as we examined above, the $m=0$ and $m=-1$ copies overlap in $k^0 \in (r_-(\vec{k}) - 2\Theta, -r_-(\vec{k}))$, and the $m=0$ and $m=-2$ copies overlap in $k^0 \in (-r_+(\vec{k}), r_+(\vec{k}) - 4\Theta)$.
We then choose corresponding solutions for the filters, and can combine them with the previous solutions to obtain, for $\Omega < |\vec{k}| < \sqrt{3/2} \Omega$:
\begin{equation}\label{eq:filtersoln2}
  H_1(k^0,\vec{k}) = H_2(-k^0,\vec{k}) =
  \begin{cases}
    \frac{\pi}{\Theta} ( 1 - e^{-i 4 \Theta \tau} )^{-1}, & \text{if $k^0 \in [-r_-(\vec{k}) + 2\Theta, r_+(\vec{k}))$} \\
    \frac{\pi}{\Theta} ( 1 - e^{-i 2 \Theta \tau} )^{-1}, & \text{if $k^0 \in (r_-(\vec{k}), -r_-(\vec{k}) + 2\Theta)$} \\
    \frac{\pi}{\Theta} ( 1 - e^{i 2 \Theta \tau} )^{-1}, & \text{if $k^0 \in (r_-(\vec{k}) - 2\Theta, -r_-(\vec{k}))$} \\
    \frac{\pi}{\Theta} ( 1 - e^{i 4 \Theta \tau} )^{-1}, & \text{if $k^0 \in (-r_+(\vec{k}), r_-(\vec{k}) - 2\Theta]$} \\
    0, & \text{otherwise}.
  \end{cases}
\end{equation}
For this solution, we require $\tau \not\in \frac{\pi}{2\Theta} \mathbb{Z}$.

Fortunately, the remaining cases are simpler.
We have already argued that for $-\Omega^2 + \Theta^2 m^2 < \vec{k}^2 < \Omega^2 + \Theta^2 m^2$, copy $|m|$ of the negative band is overlapping copy $0$ of the positive band, and copy $-|m|$ of the positive band is overlapping copy $0$ of the negative band.
The case we just considered concludes the case $|m| = 1$, and what remains is the region $|\vec{k}| \geq \sqrt{3/2} \Omega$.
Now we notice that for $|m| \geq 2$, the regions $-\Omega^2 + \Theta^2 m^2 < \vec{k}^2 < \Omega^2 + \Theta^2 m^2$ are disjoint for different $|m|$.
Therefore, if $|\vec{k}|$ is in this interval for some fixed $|m|$, then $k^0 \in (r_-(\vec{k}),r_+(\vec{k}))$ contains a subinterval where the $0$ and $|m|$ copies overlap, and the remainder of the interval consists only of copy $0$.
Regardless of how it is partitioned, we can always choose the solutions \eqref{eq:filtersolnoverlapM_1} and \eqref{eq:filtersolnoverlapM_2} with $M=|m|$ in the entire interval $k^0 \in (r_-(\vec{k}),r_+(\vec{k}))$.
Similarly, we can pick solutions \eqref{eq:filtersolnoverlapM_1} and \eqref{eq:filtersolnoverlapM_2} with $M=-|m|$ for $k^0 \in (-r_+(\vec{k}),-r_-(\vec{k}))$.
This covers all of the regions $-\Omega^2 + \Theta^2 m^2 < \vec{k}^2 < \Omega^2 + \Theta^2 m^2$.
There are also the regions in between these, namely, $\Omega^2 + \Theta^2 (|m|-1)^2 < \vec{k}^2 < -\Omega^2 + \Theta^2 m^2$ where we only have copy $0$.
Here we can choose the solutions \eqref{eq:filtersolnoverlapM_1} and \eqref{eq:filtersolnoverlapM_2} for any $M$, so we will simply extend our solutions in the region $-\Omega^2 + \Theta^2 m^2 < \vec{k}^2 < \Omega^2 + \Theta^2 m^2$ to $\Omega^2 + \Theta^2 (|m|-1)^2 \leq \vec{k}^2 < \Omega^2 + \Theta^2 m^2$, thus obtaining a partition of the entire range of $\vec{k}$.
Therefore, our filters are completely defined by \eqref{eq:filtersoln1}, \eqref{eq:filtersoln2}, and, for each $m \geq 2$, in $\Omega^2 + \Theta^2 (m-1)^2 \leq \vec{k}^2 < \Omega^2 + \Theta^2 m^2$ we choose
\begin{equation}\label{eq:filtersoln3}
  H_1(k^0,\vec{k}) = H_2(-k^0,\vec{k}) =
  \begin{cases}
    \frac{\pi}{\Theta} ( 1 - e^{-i 2 \Theta \tau m} )^{-1}, & \text{if $k^0 \in (r_-(\vec{k}),r_+(\vec{k}))$} \\
    \frac{\pi}{\Theta} ( 1 - e^{i 2 \Theta \tau m} )^{-1}, & \text{if $k^0 \in (-r_+(\vec{k}),-r_-(\vec{k}))$} \\
    0, & \text{otherwise}.
  \end{cases}
\end{equation}
Note that the combined solution requires that $\tau \not\in \frac{\pi}{m \Theta} \mathbb{Z}$ for all $m \in \mathbb{Z} \setminus \{ 0 \}$.
This is equivalent to requiring $\tau \in \frac{\pi}{\Theta} ( \mathbb{R} \setminus \mathbb{Q} )$.

We have thus demonstrated that $|\phi) = \sum_{i=1}^2 H_i S_i |\phi)$ for any bandlimited $\phi$ with the above choices for $H_i(k)$.
Therefore, for our lattice $\{ t_n^{(1)} = n\pi/\Theta \}_{n \in \mathbb{Z}} \cup \{ t_n^{(2)} = n\pi/\Theta + \tau \}_{n \in \mathbb{Z}}$, with $\Theta = \Omega/\sqrt{2}$ and $\tau \in \tfrac{\pi}{\Theta} ( \mathbb{R} \setminus \mathbb{Q} )$, we have the reconstruction formula:
\begin{align}
  \phi(x) &\equiv (x|\phi) \\
  &= (x| \sum_{i=1}^2 H_i S_i |\phi) \\
  &= \sum_{n \in \mathbb{Z}} \int d^N\vec{x}' \left[ (x| H_1 |t_n^{(1)},\vec{x}') \phi(t_n^{(1)},\vec{x}') + (x| H_2 |t_n^{(2)},\vec{x}') \phi(t_n^{(2)},\vec{x}') \right]. \label{eq:minkowski_reconstruction}
\end{align}
We have thus succeeded in demonstrating that a reconstruction of the form \eqref{eq:Minkowksi_reconstruction_allmode} for an arbitrary Minkowski-bandlimited function $\phi$ can be achieved with a single temporal lattice.
In this formula, we have two different forms of reconstruction kernel: $(x| H_1 |t_n^{(1)},\vec{x}')$ for sublattice $\{ t_n^{(1)} \}_{n \in \mathbb{Z}}$ and $(x| H_2 |t_n^{(2)},\vec{x}')$ for sublattice $\{ t_n^{(2)} \}_{n \in \mathbb{Z}}$.
Note that since $H_2(k^0,\vec{k}) = H_1(-k^0,\vec{k})$, then $(x|H_2|x')$ is simply a time-reflected $(x|H_1|x')$.

Let us write $K(t-t';\vec{x}-\vec{x}') := (t,\vec{x}|H_1|t',\vec{x}')$, and thus $(t,\vec{x}|H_2|t',\vec{x}') = K(t'-t;\vec{x}-\vec{x}')$.
Since the filters are piecewise constant in Fourier space, we can easily evaluate the spatial Fourier transform of $K$.
The result is, for $|\vec{k}| \leq \Omega$,
\begin{equation}
  \tilde{K}_{\vec{k}}(t) = \frac{1}{2 \Theta t \sin(\Theta \tau)} \left[ \cos( r_+(\vec{k}) t - \Theta \tau ) - \cos(\Theta \tau) \right],
\end{equation}
for $\Omega < |\vec{k}| < \sqrt{3/2} \Omega$,
\begin{align}
  \tilde{K}_{\vec{k}}(t) &= \frac{1}{2 \Theta t \sin(\Theta \tau)} \left[ \cos[ ( -r_-(\vec{k}) + 2 \Theta ) t - \Theta \tau ] - \cos[ r_-(\vec{k}) t - \Theta \tau ] \right] \nonumber \\
  &\qquad + \frac{1}{2 \Theta t \sin(2 \Theta \tau)} \left[ \cos[ r_+(\vec{k}) t - 2 \Theta \tau ] - \cos[ ( -r_-(\vec{k}) + 2 \Theta ) t - 2 \Theta \tau ] \right],
\end{align}
and for each $m \geq 2$ and $\Omega^2 + \Theta^2 (m-1)^2 \leq \vec{k}^2 < \Omega^2 + \Theta^2 m^2$,
\begin{equation}
  \tilde{K}_{\vec{k}}(t) = \frac{1}{2 \Theta t \sin(m \Theta \tau)} \left[ \cos( r_+(\vec{k}) t - m \Theta \tau ) - \cos( r_-(\vec{k}) t - m \Theta \tau ) \right].
\end{equation}
The inverse spatial Fourier transform of $\tilde{K}_{\vec{k}}(t)$ will, in principle, give $K(t,\vec{x})$, although we will not attempt to calculate this here.

Note that these are not the unique reconstruction kernels for this lattice.
This is because there is freedom in the choice of filters both inside and outside the bandlimited region of Fourier space.
The freedom inside the region is from the one-dimensional space of solutions for the two filters at each point where we only have support of the $m=0$ copy.
The freedom outside is because the filters can be chosen arbitrarily in a region where none of the copies have support (or could even be nonzero in regions where one of the copies $m \neq 0$ has support, provided the filters are chosen so that this term does not appear in the sum of \eqref{eq:minkowskisecondordercopies}).
The latter type of freedom is analogous to that occurring when oversampling for the one-dimensional Shannon sampling we discussed in Subsection~\ref{subsec:shannon_formula}.
The former kind, where one has a space of solutions for the filters within the bandlimited region, only occurs in higher-order sampling.

Indeed, these reconstruction kernels are not exactly the same as those previously stated in \cite{KempfChatwindaviesMartin2013}, however the difference can be accounted for with this freedom we have identified in the choice of the filters.
In \cite{KempfChatwindaviesMartin2013}, for a fixed $|\vec{k}| \leq \Omega$, instead of restricting attention to the interval $k^0 \in (-r_+(\vec{k}),r_+(\vec{k}))$ on which copy $0$ is supported, the authors looked at the larger ($\vec{k}$-independent) interval $k^0 \in (-2\Theta,2\Theta)$.
This contains the previous interval, and the only copies supported in this interval are again $m= -1, 0, \text{ and } 1$.
The $m=0$ term in the sum of \eqref{eq:minkowskisecondordercopies} can similarly be isolated by choosing the filter solutions \eqref{eq:filtersolnoverlapM_1} and \eqref{eq:filtersolnoverlapM_2} with $M=1$ in $k^0 \in [0,2\Theta)$ and $M=-1$ in $k^0 \in (-2\Theta,0)$.
This differs from the solution presented here since the filters are now nonzero outside of the support of the $0^{th}$ copy, albeit in a way where \eqref{eq:minkowskisecondordercopies} is still satisfied.
Similarly, for a fixed $|\vec{k}| > \Omega$, they consider the interval $k^0 \in (r_-(\vec{k}), r_-(\vec{k}) + 2\Theta)$, which contains only the $0^{th}$ copy of the positive band, and copies $M_{\vec{k}} := \lfloor r_-(\vec{k})/\Theta + 1 \rfloor$ and $M_{\vec{k}} + 1$ of the extended negative band $(-r_-(\vec{k}) - 2\Theta, -r_-(\vec{k}))$ (unless the quantity $r_-(\vec{k})/\Theta + 1$ is an integer, in which case we only have the $M_{\vec{k}}$ copy of the negative band).
Equation \eqref{eq:minkowskisecondordercopies} is then satisfied by choosing solutions \eqref{eq:filtersolnoverlapM_1} and \eqref{eq:filtersolnoverlapM_2} with $M=M_{\vec{k}}$ in $k^0 \in ( r_-(\vec{k}), -r_-(\vec{k}) + 2 \Theta M_{\vec{k}} ]$ and $M=M_{\vec{k}}+1$ in $k^0 \in ( -r_-(\vec{k}) + 2 \Theta M_{\vec{k}}, r_-(\vec{k}) + 2 \Theta )$.
The analysis for the $0^{th}$ copy of the negative band is similar.
Again, these filters will have support outside of the support of copy $0$.
Spelling all this out gives the solution stated in \cite{KempfChatwindaviesMartin2013}, after replacing our $M_{\vec{k}}$ with their $m_{\vec{k}}$ by $M_{\vec{k}} = 2 m_{\vec{k}} - 1$.\footnote{
For the reconstruction formula stated in \cite{KempfChatwindaviesMartin2013}, which is taken from \cite{Linden1959}, $m_{\vec{k}}$ is defined to be $\lceil r_-(\vec{k}) / 2\Theta \rceil$, in our notation.
Our identification $M_{\vec{k}} = 2 m_{\vec{k}} - 1$ would lead one to suspect that $M_{\vec{k}}$ is always odd.
However, the figures drawn in \cite{Linden1959} implicitly assume that $r_-(\vec{k})/2\Theta + 1/2 < \lceil r_-(\vec{k})/2\Theta \rceil$, which is of course not always true.
This is fine for purposes of illustration, but the failure to acknowledge the remaining case in designing the filters leads one to the erroneous conclusion that the copies of the negative band which overlap the $0^{th}$ copy of the positive band are always $2 m_{\vec{k}} - 1$ and $2 m_{\vec{k}}$.
However, in situations where $r_-(\vec{k})/2\Theta  + 1/2 > \lceil r_-(\vec{k})/2\Theta \rceil$, which for us will always occur for some values of $\vec{k}$, the overlap is between copies $2 m_{\vec{k}}$ and $2 m_{\vec{k}} + 1$.
Therefore, the reconstruction formula stated in \cite{Linden1959} and \cite{KempfChatwindaviesMartin2013} is not quite correct in these cases.
Our expressions using $M_{\vec{k}} = \lfloor r_-(\vec{k})/\Theta \rfloor + 1$ do not make any assumptions such as this, and agrees with the analysis of \cite{Kohlenberg1953}.
}
In particular, the spatial Fourier transform of the reconstruction kernel one obtains is, for $|\vec{k}| \leq \Omega$,
\begin{equation}
  \tilde{K}_{\vec{k}}(t) = \frac{1}{2 \Theta t \sin(\Theta \tau)} \left[ \cos( 2 \Theta t - \Theta \tau ) - \cos(\Theta \tau) \right],
\end{equation}
and for $|\vec{k}| > \Omega$,
\begin{align}
  \tilde{K}_{\vec{k}}(t) &= \frac{1}{2 \Theta t \sin( (M_{\vec{k}} + 1) \Theta \tau )} \left[ \cos[ (r_-(\vec{k}) + 2\Theta) t - (M_{\vec{k}} + 1) \Theta \tau ] \right. \nonumber \\
  & \qquad \qquad \qquad \qquad \qquad \qquad \qquad \left. - \cos[ (-r_-(\vec{k}) + 2 \Theta M_{\vec{k}}) t - (M_{\vec{k}} + 1) \Theta \tau ] \right] \nonumber \\
  & \qquad + \frac{1}{2 \Theta t \sin( M_{\vec{k}} \Theta \tau )} \left[ \cos[ (-r_-(\vec{k}) + 2 \Theta M_{\vec{k}}) t - M_{\vec{k}} \Theta \tau ] - \cos[ r_-(\vec{k}) t - M_{\vec{k}} \Theta \tau ] \right].
\end{align}
Therefore, we see that the freedom that we identified for the filters can yield different forms for the reconstruction kernels, even for the same set of samples.
They are, however, equivalent.

In our derivation of the Minkowski temporal sampling formula, we found that the two sublattices of sample points in time must be shifted from one another by some $\tau \in \frac{\pi}{\Omega/\sqrt{2}} ( \mathbb{R} \setminus \mathbb{Q} )$.
This requirement was not previously identified, and it suggests that there might be an instability in the reconstruction of Minkowski-bandlimited functions from this set of samples, since, roughly speaking, $\tau$ must be tuned with infinite precision.
Of course, this would be a significant practical limitation, but it would be interesting to know whether the sampling is stable despite this strict requirement.
If the sampling is not stable, then the reconstruction from samples would not be continuous, and this could have important consequences if the reconstruction formula is used in, for example, a path integral to replace fields on spacetime with their samples.
However, this, along with other stability considerations will be studied elsewhere.

In this section, we have seen that despite the fact that Minkowski-bandlimited functions require an infinite density of sample points in spacetime, they do exhibit the property that they can be reconstructed from knowledge of the functions at a discrete set of points in time.
Recall that the first step we took in developing this was a temporal-spatial splitting of the bandlimited region in Fourier space.
Although this split assisted in highlighting the reconstruction properties of these bandlimited functions, it has obscured the Poincar\'e-invariant nature of the space of functions, particularly under Lorentz boosts.
Of course, the bandlimitation condition $|k^2| < \Omega^2$ is Lorentz-invariant, and thus so is the corresponding space of bandlimited functions.
However, taking samples on constant-time hypersurfaces would appear to identify a preferred frame, and indeed the reconstruction formula \eqref{eq:minkowski_reconstruction} is not manifestly covariant.
How can this be consistent with the symmetry?
Understanding the covariance of the sampling and reconstruction we have outlined here will be the subject of the next section.

\section{Sampling and symmetries}
\label{sec:sampling_symmetries}

How can the sampling and reconstruction for the Minkowski-bandlimited functions be consistent with the Poincar\'e-invariance of the function space?
An intuitive argument was given in \cite{Kempf2004a,KempfChatwindaviesMartin2013} for how this works.
Consider a fixed spatial mode $\vec{k}$.
This has a finite temporal bandwidth, and can be sampled from some discrete set of points in time, $\{ t_{\vec{k},n} \}_{n \in \mathbb{Z}}$, spaced with a density proportional to the bandwidth.
Under a boost, the time coordinate undergoes dilation, and so this lattice appears less dense in the new coordinates.
However, the spatial frequency $\vec{k}$ also changes under the boost to become a larger spatial frequency $\vec{k}'$ due to length contraction.
The temporal bandwidth of a larger fixed spatial mode $\vec{k}'$ is smaller, thus for sampling this mode requires a temporal lattice which is less dense, consistent with the time dilation of the original temporal lattice.
This is how the temporal lattice with the smaller density can still suffice to sample and reconstruct this mode in the new frame.

However, upon further scrutiny, we see that there are issues with this kind of reasoning.
First, it is not clear how this would apply when sampling all of the modes from the same temporal lattice, since the lattice would become less dense for all the modes, yet the function space should map into itself and hence require a lattice of the original spacing.
Second, given two spatial modes $\vec{k}$ and $\vec{k}'$ with $|\vec{k}'| > |\vec{k}|$, the temporal bandwidth of mode $\vec{k}'$ is not always smaller than that of $\vec{k}$.
It is true if $|\vec{k}'| > |\vec{k}| > \Omega$, but not for $|\vec{k}| < |\vec{k}'| < \Omega$.
Recall we identified the maximum temporal bandwidth to be $2\sqrt{2}\Omega$ at $|\vec{k}| = \Omega$, so, starting with $\vec{k} = 0$, the bandwidth first increases before decreasing beyond the Planck scale.
The main issue with this type of argument, however, stems from the fact that a fixed spatial mode does not map into a fixed spatial mode under a boost.
A spatial mode consists of a single $\vec{k}$ and a range of values of $k^0$.
Under a boost, the range of values of the temporal component will mix with the spatial component, and so the image will consist of a line spanning a range of values of $\vec{k}'$.
Hence, it is not simply a fixed spatial mode, and the development of the reconstruction formula in the previous section does not imply that after a boost this mode can be reconstructed from a discrete set of constant-time hypersurfaces in the new coordinates.

How then does the symmetry manifest itself in the sampling and reconstruction?
Let us first examine the simple one-dimensional case of sampling functions in $L^2(\mathbb{R})$ bandlimited to $|k| < \Omega$.
This bandlimited subspace is translation-invariant, provided that these functions vanish on the boundary of the interval $k \in [-\Omega,\Omega]$.
This is because a translation by $a \in \mathbb{R}$ acts in Fourier space by $\tilde{\phi}(k) \mapsto e^{-i a k} \tilde{\phi}(k)$.
Such an operation does not change the support of $\tilde{\phi}(k)$, and if $\tilde{\phi}(k)$ vanishes at the boundary of the bandlimited interval, then its image under the translation will as well.
Above we showed that any function in this bandlimited space can be reconstructed from its values on the lattice $\{ x_n = n\pi/\Theta \}_{n \in \mathbb{Z}}$, as long as $\Theta \geq \Omega$.
This was accomplished by demonstrating that the sampling operator, $S = \sum_{n \in \mathbb{Z}} |x_n)(x_n|$, and the filter, $H = \frac{\pi}{\Theta} \int_{-\Omega}^\Omega \frac{dk}{2\pi} |k)(k|$, satisfy $H S |\phi) = |\phi)$ for any bandlimited $\phi$.
How is the translation-symmetry of the function space consistent with the fact that these functions can be represented on a discrete sampling lattice?
Well, suppose instead of the lattice $\{ x_n = n\pi/\Theta \}_{n \in \mathbb{Z}}$, we chose a translated version of this lattice, $\{ x_n^{(\alpha)} = (n + \alpha) \pi / \Theta \}_{n \in \mathbb{Z}}$ for some fixed $\alpha \in [0,1)$.
In this case, our constraint $|\phi) \stackrel{!}{=} H S |\phi)$ in Fourier space becomes:
\begin{equation}
  \tilde{\phi}(k) \stackrel{!}{=} \frac{\Theta}{\pi} H(k) \sum_{m \in \mathbb{Z}} e^{- 2 \pi i \alpha m} \tilde{\phi}(k - 2 \Theta m).
\end{equation}
Provided $\Theta \geq \Omega$, we can solve this using the same filter as before, and obtain the reconstruction formula:
\begin{equation}
  \phi(x) = \sum_{n \in \mathbb{Z}} \sinc [ \Omega ( x - x_n^{(\alpha)} ) ] \phi(x_n^{(\alpha)}).
\end{equation}
Therefore, we see that we have a family of translated lattices which can be used in the sampling formula for the reconstruction.
Also, none of these lattices are preferred, as no choice of $\alpha$ is distinguished.\footnote{It is interesting to examine the importance of the assumption of Dirichlet boundary conditions for the Fourier space interval.
Suppose we were to consider imposing the more general condition of periodicity on $k \in [-\Omega,\Omega]$ up to a phase: $\tilde{\phi}(\Omega) = e^{- 2 \pi i \varphi} \tilde{\phi}(-\Omega)$.
One can show that this space of functions can only be reconstructed from its values on the lattice $\{ x_n^{(\alpha)} = (n + \alpha) \pi / \Omega \}_{n \in \mathbb{Z}}$ if $\alpha = \varphi$, since
\begin{equation}
  |\phi) = \sum_{n \in \mathbb{Z}} H |x_n^{(\alpha)})(x_n^{(\alpha)}|\phi) \quad \implies \quad \tilde{\phi}(k) = \frac{\pi}{\Omega} \sum_{n \in \mathbb{Z}} e^{-i k x_n^{(\alpha)}} \phi(x_n^{(\alpha)}) \quad \implies \quad \tilde{\phi}(\Omega) = e^{-2 \pi i \alpha} \tilde{\phi}(-\Omega),
\end{equation}
where $H = \frac{\pi}{\Omega} \int_{-\Omega}^\Omega \frac{dk}{2\pi} |k)(k|$.
This function space is also only invariant under translations $\tilde{\phi}(k) \mapsto e^{-i a k} \tilde{\phi}(k)$ if $a \in \frac{\pi}{\Omega} \mathbb{Z}$, which are a discrete subset of translations mapping this lattice into itself.
Therefore, we have a restricted symmetry group which is reflected in there being a single lattice at the minimal density from which the functions can be reconstructed.
This is the situation which occurs in crystals, for example.}

The existence of a family of sampling lattices in the translation-symmetric case can intuitively be seen as the manifestation of the symmetry.
The discreteness of the sampling lattice at first appears incongruous with the symmetry since the lattice does not generally map into itself under a symmetry transformation.
The lattice would therefore seem to pick out a preferred frame of reference.
However, the fact that there is a family of sampling lattices from which one can reconstruct dispels this tension, since it demonstrates that an observer in any translated frame could have constructed a similar sampling formula in their frame.
Thus, in the end none of the frames are preferred.
That there is no preferred origin in the Shannon sampling theorem was already understood even in the earliest sampling theory literature \cite{Whittaker1915}, but one would expect that the above observations should apply generally.
For example, in the Minkowski case, we could have began our development of the reconstruction formula from any inertial frame.
The formula which one arrives at in one inertial frame could then be used in any other inertial frame, yielding a family of sampling lattices and reconstruction formulas related by Poincar\'e transformations.
Although it is intuitively clear this can be done, here we would like to explicitly elaborate upon how this works in general.
In particular, our aim is the clarify the relationship between spacetime symmetries and the existence of families of sampling lattices, as well as to write out explicitly the form of the reconstruction kernel for any choice of lattice in the family.
We will first examine a general scenario before returning to our case of Minkowski bandlimitation.

\subsection{General considerations}
\label{subsec:genl_considerations}

Consider the Hilbert space $L^2(X)$ of square-integrable functions over some space $X$.
For each element $x \in X$, we have a corresponding $|x)$ which gives the representation of an element $|\phi) \in L^2(X)$ by $\phi(x) \equiv (x|\phi)$.
Now suppose we have some symmetry group, $G$, with each $g \in G$ acting on $X$ as an isomorphism $x \mapsto g(x)$, and which is unitarily represented on $L^2(X)$ by $U(g)|x) = |g(x))$.\footnote{
Note that here we assume that the measure on $X$ used to define the inner product on $L^2(X)$ is invariant under $G$.
This is of course true for the cases we have in mind in this paper.
}
Now, let $\mathcal{B}$ be a subspace of $L^2(X)$ such that we have a reconstruction formula of the general form:
\begin{equation}
  \phi(x) = \sum_{n \in \mathbb{Z}} K_n(x) \phi(x_n) \qquad \forall |\phi) \in \mathcal{B},
\end{equation}
for some discrete set $\{ x_n \}_{n \in \mathbb{Z}} \subset X$ and some set of functions $\{ K_n(x) \}_{n \in \mathbb{Z}}$.
Now we want to see what the assumption of the invariance of $\mathcal{B}$ under the action of $U(G)$ means for the reconstruction.
Of course, we have in mind cases where $U(G)$ does not simply map the lattice $\{ x_n \}_{n \in \mathbb{Z}}$ into itself.

It will be helpful to introduce some notation to label different coordinate frames.
Let us write $|x)_{\mathcal{O}}$ as the element corresponding to the point labelled by the coordinate $x$ in the frame $\mathcal{O}$.
We can then view the action $U(g)|x)_{\mathcal{O}} = |g(x))_{\mathcal{O}}$ as an active transformation moving the point labelled by $x$ to that labelled by $g(x)$ in the frame $\mathcal{O}$.
We could also pick some element $g \in G$, and construct a new frame $\mathcal{O}'$ by taking the point labelled by $x$ in the frame $\mathcal{O}$ to a new label $g(x)$.
Since these are physically the same point, we can express this passive coordinate transformation by $|g(x))_{\mathcal{O}'} = |x)_{\mathcal{O}}$.
This new frame then inherits the action of the symmetry group, $U(g)|x)_{\mathcal{O}'} = U(g)|g^{-1}(x))_{\mathcal{O}} = |x)_{\mathcal{O}} = |g(x))_{\mathcal{O}'}$.
We also see that active and passive transformations can always be identified, since we can write $|x)_{\mathcal{O}'} = U(g)^\dagger |x)_{\mathcal{O}}$.

For some abstract element $|\phi)$ in the Hilbert space, we will write the function representing this element in the frame $\mathcal{O}$ as $\phi^{\mathcal{O}}(x) \equiv {}_{\mathcal{O}} (x|\phi)$.
We can relate the representations of $|\phi)$ in different frames by $\phi^{\mathcal{O'}}(x) = {}_{\mathcal{O}'} (x|\phi) = {}_{\mathcal{O}} (g^{-1}(x)|\phi) = \phi^{\mathcal{O}}(g^{-1}(x))$, demonstrating that $\phi^{\mathcal{O}'} = \phi^{\mathcal{O}} \circ g^{-1}$ (i.e., $\phi^{\mathcal{O}'}$ is the pullback of $\phi^{\mathcal{O}}$ under $g^{-1}$).

Now suppose we have a reconstruction formula for elements in the subspace $\mathcal{B}$ which was developed in the frame $\mathcal{O}$, using a sampling lattice $\{ x_n \}_{n \in \mathbb{Z}}$:
\begin{equation}
  \phi^{\mathcal{O}}(x) = \sum_{n \in \mathbb{Z}} K_n^{\mathcal{O}}(x) \phi^{\mathcal{O}}(x_n).
\end{equation}
We can also write this abstractly as:
\begin{equation}
  |\phi) = \sum_{n \in \mathbb{Z}} |K_n) \left._{\mathcal{O}} (x_n|\phi) \right..
\end{equation}
This can then be expressed in a different coordinate frame $\mathcal{O}'$ as:
\begin{equation}\label{eq:genlreconstruction_covariance_0}
  \phi^{\mathcal{O}'}(x) = \sum_{n \in \mathbb{Z}} K_n^{\mathcal{O}'}(x) \phi^{\mathcal{O}'}(g(x_n)).
\end{equation}
Note that the sampling lattice in this frame is simply the image of the original lattice under the coordinate transformation $g$.
This is not a different sampling of the function, just the same reconstruction formula expressed in a different coordinate frame.
It simply follows from the assertion that our functions transform as scalars under coordinate transformations.

If the subspace $\mathcal{B}$ is invariant under the symmetry group then we get more.
Suppose $U(g)|\phi) \in \mathcal{B}$ for every $|\phi) \in \mathcal{B}$ and every $g \in G$.
Then for each $g \in G$, we get a reconstruction formula for $U(g)|\phi)$.
Therefore,
\begin{align}
  U(g) |\phi) &= \sum_{n \in \mathbb{Z}} |K_n) \left._{\mathcal{O}} (x_n| U(g) |\phi) \right. \\
  {}_{\mathcal{O}} (x| U(g) |\phi) &= \sum_{n \in \mathbb{Z}} {}_{\mathcal{O}} (x|K_n) \left._{{\mathcal{O}}} (x_n| U(g) |\phi) \right. \\
  \phi^{\mathcal{O}_g}(x) &= \sum_{n \in \mathbb{Z}} K_n^{\mathcal{O}}(x) \phi^{\mathcal{O}_g}(x_n),
\end{align}
where by $\mathcal{O}_g$ we mean the frame obtained from $\mathcal{O}$ by the coordinate transformation $x \mapsto g(x)$.
Therefore, we see that we can reconstruct from the sampling lattice $\{ x_n \}_{n \in \mathbb{Z}}$, but now in the frame $\mathcal{O}_g$, using the same functions $K_n^{\mathcal{O}}(x)$ which were used in the original frame $\mathcal{O}$.
Although these are the same functions of $x$, now $x$ is more readily interpreted in the above equation as the coordinate of a point in the $\mathcal{O}_g$ frame.
These functions can equivalently be associated with the Hilbert space elements $U(g)^\dagger |K_n)$, since ${}_{\mathcal{O}_g} (x| U(g)^\dagger |K_n) = {}_\mathcal{O} (x|K_n) = K_n^{\mathcal{O}}(x)$.
Therefore, $K_n^{\mathcal{O}}(x)$ should not be thought of as the same function on space as before, but just takes the same form in the new frame.
The equation above then reflects the statement that, if $\mathcal{B}$ is invariant under $U(G)$, then one can build the same reconstruction formula starting in any frame related by the symmetry group transformations.
This is clearly different from \eqref{eq:genlreconstruction_covariance_0}, which is merely showing that the reconstruction formula from one frame can be expressed in any other frame.

How can we see the family of sampling lattices?
We can equivalently express the abstract reconstruction formula for $U(g)|\phi)$ as:
\begin{align}
  |\phi) &= \sum_{n \in \mathbb{Z}} U(g)^\dagger |K_n) \left._{\mathcal{O}} (x_n| U(g) |\phi) \right. \\
  {}_{\mathcal{O}} (x|\phi) &= \sum_{n \in \mathbb{Z}} {}_{\mathcal{O}} (x| U(g)^\dagger |K_n) \left._{\mathcal{O}} (x_n| U(g) |\phi) \right. \\
  \phi^{\mathcal{O}}(x) &= \sum_{n \in \mathbb{Z}} K_n^{\mathcal{O}}(g(x)) \phi^{\mathcal{O}}(g^{-1}(x_n)).
\end{align}
Therefore, we see that in the same frame $\mathcal{O}$ we can sample from the new lattice $\{ g^{-1}(x_n) \}_{n \in \mathbb{Z}}$ using the reconstruction kernels $\{ (K_n^{\mathcal{O}} \circ g)(x) \}_{n \in \mathbb{Z}}$.
Since this applies for any $g \in G$, we see that we obtain a family of lattices related by the symmetry transformations.

Therefore, to summarize, if in some frame $\mathcal{O}$ we have
\begin{equation}\label{eq:genlreconstruction}
  \phi^\mathcal{O}(x) = \sum_{n \in \mathbb{Z}} K_n(x) \phi^{\mathcal{O}}(x_n)
\end{equation}
where we simply write $K_n(x) \equiv {}_{\mathcal{O}} (x|K_n)$, and if $U(G) \mathcal{B} \subseteq \mathcal{B}$, then
\begin{equation}\label{eq:genlreconstruction_family}
  \phi^{\mathcal{O}}(x) = \sum_{n \in \mathbb{Z}} K_n(g^{-1}(x)) \phi^{\mathcal{O}}(g(x_n)) \qquad \forall g \in G.
\end{equation}
This demonstrates that, given one sampling lattice $\{ x_n \}_{n \in \mathbb{Z}}$, the symmetry of $\mathcal{B}$ implies that there is always a family of sampling lattices related by the symmetry transformations.
Further, the reconstruction formula for all of these lattices is given by \eqref{eq:genlreconstruction_family}, in terms of the reconstruction kernels $\{ K_n \}_{n \in \mathbb{Z}}$ for the original lattice.
The reconstruction formula above is equivalent to the statement that
\begin{equation}\label{eq:genlreconstruction_otherframes}
  \phi^{\mathcal{O}_g}(x) = \sum_{n \in \mathbb{Z}} K_n(x) \phi^{\mathcal{O}_g}(x_n)
\end{equation}
for any frame $\mathcal{O}_g$ related to $\mathcal{O}$ by the transformation $x \mapsto g(x)$.
This is stronger than simply the statement that the original formula can be used in different frames, which applies even if $\mathcal{B}$ is not invariant under $U(G)$:
\begin{equation}\label{eq:genlreconstruction_covariance}
  \phi^{\mathcal{O}_g}(x) = \sum_{n \in \mathbb{Z}} K_n(g^{-1}(x)) \phi^{\mathcal{O}_g}(g(x_n)).
\end{equation}

We can also generate a kind of converse statement to the one above.
One may ask whether \eqref{eq:genlreconstruction_family} holding for all $|\phi)$ in some space $\mathcal{B}$ implies that $\mathcal{B}$ is invariant under $U(G)$.
Not exactly, since suppose we took the one-dimensional space spanned by some particular $|\phi)$ for which \eqref{eq:genlreconstruction_family} holds.
It is not guaranteed that $U(g)|\phi) \in \text{span}_{\mathbb{C}} \{ |\phi) \}$ for all $g \in G$.
However, if \eqref{eq:genlreconstruction_family} holds for $|\phi)$, then it automatically holds for any $U(g)|\phi)$.
It will therefore also hold for any finite linear combination of these elements, which we write as the space $\text{span}_{\mathbb{C}} \{ U(g) |\phi) \}_{g \in G}$.
Note that this space is invariant under $U(G)$ by construction.
We can extend these statements to the closure of this space as well.
Since $\sum_{n \in \mathbb{Z}} U(g) |K_n) \left._{\mathcal{O}}(x_n| U(g)^\dagger \right.$ is assumed to act as the identity on this space (for each $g \in G$), and $U(g)$ is unitary, then these operators are bounded and can be extended by continuity to the closure of this space, $\overline{\text{span}}_{\mathbb{C}} \{ U(g) |\phi) \}_{g \in G}$.
The operator $\sum_{n \in \mathbb{Z}} U(g) |K_n) \left._{\mathcal{O}}(x_n| U(g)^\dagger \right.$ again acts as the identity on the closure.
Thus, \eqref{eq:genlreconstruction_family} holds on $\overline{\text{span}}_{\mathbb{C}} \{ U(g) |\phi) \}_{g \in G}$ and this space is invariant under $U(G)$.
One can apply the same arguments to an arbitrary space $\mathcal{B}$ for which \eqref{eq:genlreconstruction_family} holds.
Therefore, we can conclude that if \eqref{eq:genlreconstruction_family} holds for every element in a space $\mathcal{B}$, then it can always be extended to hold on $\overline{\text{span}}_{\mathbb{C}} \{ U(g) |\phi) : g \in G, |\phi) \in \mathcal{B} \}$, which is invariant under $U(G)$.
This is also the smallest Hilbert space containing $\mathcal{B}$ with these properties.

Therefore, we see on general grounds how the symmetry of a function space with a sampling formula is equivalent to the existence of a family of sampling lattices related by the symmetry transformations.
Note that the assumed form for the reconstruction formula \eqref{eq:genlreconstruction} was quite general, demonstrating that this feature is independent of the method used to obtain the reconstruction formula.

A particular family of lattices of the form $\{ g(x_n) \}_{n \in \mathbb{Z}}$ will not generally exhaust the full set of possible sampling lattices.
For example, in the one-dimensional translation-invariant case, lattices of the form $\{ x_n = n\pi/\Theta \}_{n \in \mathbb{Z}}$ with different values of $\Theta \geq \Omega$ are not related by translations.
Generally, one would expect there to be many lattices of sufficiently large Beurling density which could be used for sampling and reconstruction.
The point here is that every sampling lattice lives in some family of lattices related by the symmetry, and each lattice in this family will have sufficiently large density.
Further, given the reconstruction formula for one, we automatically get those for any other in the corresponding family.

Notice that in the one-dimensional case we considered above, the reconstruction kernel only depends on the index $n$ through $x_n$, and so we can write it as $K_n(x) = K(x,x_n)$.
Also, recall that the reconstruction kernel we used for the translated sampling lattices was of the form $K(x,g(x_n))$, i.e., we simply use the same kernel and plug in the new lattice points.
This does not look exactly like the form $K(g^{-1}(x),x_n)$ specified by \eqref{eq:genlreconstruction_family}.
However, in the one-dimensional case these are equivalent since the kernel has the special property $K(g(x),g(x_n)) = K(x,x_n)$.
This only occurs because the reconstruction kernel is given by the matrix elements of a filter, $K(x,x_n) = (x|H|x_n)$, which is in turn proportional to the projection operator onto the bandlimited subspace, $H = \frac{\pi}{\Theta} \int_{-\Omega}^\Omega \frac{dk}{2\pi} |k)(k|$.
Since the bandlimited subspace is invariant under the symmetry group, then the projection operator onto this space must commute with the symmetry transformation operators.
The filter $H$ therefore commutes with the symmetry operator $U(g)$, and thus yields the above property for the reconstruction kernel.
The feature $K(g^{-1}(x),x_n) = K(x,g(x_n))$ however only arises in special cases.
For example, the filters we constructed for the Minkowski temporal sampling are not invariant under Lorentz boosts (i.e., $U(\Lambda)^\dagger H_i U(\Lambda) \neq H_i$).
In general, one must simply use the form presented in \eqref{eq:genlreconstruction_family}.

Note that the analysis we performed here did not uncover any particular properties exhibited by the functions $\{ K_n(x) \}_{n \in \mathbb{Z}}$, such as $K(g(x),g(x_n)) = K(x,x_n)$ occurring in the one-dimensional case.
Of course, generally these reconstruction kernels have to be carefully constructed (up to some freedom) in order to reproduce a function from its samples on a particular lattice.
However, it is interesting to note that with the general reasoning we used in this section, we were able to establish how the symmetry manifests itself in the sampling formula without making any further assumptions about the form of these functions.

This concludes our general discussion of how symmetries manifest themselves in sampling formulas, namely as a family of lattices related by the symmetry transformations from which one can reconstruct a bandlimited function.
We further showed explicitly how a bandlimited function can be reconstructed from any lattice in this family, given the reconstruction formula for one of the lattices.

\subsection{Lorentz symmetry of Minkowski temporal sampling}

The formulas presented in the general discussion above do not immediately apply to the Minkowski temporal sampling formula, since the samples consist of functions on constant-time hypersurfaces rather than numbers at points.
However, the same ideas can be extended without any difficulty.
We present it here for completeness.
We will focus only on Lorentz transformations since the analysis of translations is similar to the one-dimensional scenario.

Using similar notation as above for distinguishing between different frames, we found that in some arbitrary inertial frame $\mathcal{O}$, we could pick
\begin{equation}
  t_n^{(1)} = \frac{n \pi}{\Omega/\sqrt{2}}, \quad t_n^{(2)} = \frac{n \pi}{\Omega/\sqrt{2}} + \tau, \qquad (n \in \mathbb{Z}),
\end{equation}
for some fixed $\tau \in \frac{\pi}{\Omega/\sqrt{2}} ( \mathbb{R} \setminus \mathbb{Q} )$, with corresponding sampling operators,
\begin{equation}
  S_i = \sum_{n \in \mathbb{Z}} \int d^N\vec{x} \left. |t_n^{(i)},\vec{x})_{\mathcal{O}} \right. {}_{\mathcal{O}} (t_n^{(i)},\vec{x}|,
\end{equation}
and some appropriately chosen filters,
\begin{equation}
  H_i = \int \frac{d^{N+1}k}{(2\pi)^{N+1}} H_i(k) \left. |k)_{\mathcal{O}} \right. {}_{\mathcal{O}} (k|,
\end{equation}
so that $|\phi) = ( H_1 S_1 + H_2 S_2 ) |\phi)$.
We can also write $|\phi) = ( H_1 S_1 + H_2 S_2 ) |\phi)$ as:
\begin{equation}
  \phi^{\mathcal{O}}(x) = \sum_{n \in \mathbb{Z}} \int d^N\vec{y} \left[ K_1 (x; t_n^{(1)},\vec{y}) \hspace{1mm} \phi^{\mathcal{O}} ( t_n^{(1)},\vec{y} ) + K_2 (x; t_n^{(2)},\vec{y}) \hspace{1mm} \phi^{\mathcal{O}} (t_n^{(2)},\vec{y}) \right]
\end{equation}
where $K_i(x;y) \equiv {}_{\mathcal{O}} (x| H_i |y)_{\mathcal{O}}$.
Now, even if the bandlimited function space is not invariant under Lorentz transformations, we can always express this formula in another inertial frame, under $x \mapsto x' = \Lambda x$, as:
\begin{align}
  (\phi^{\mathcal{O}} \circ \Lambda^{-1})(x') &= \sum_{n \in \mathbb{Z}} \int d^N\vec{y} \left[ K_1 ( \Lambda^{-1} x' ; t_n^{(1)},\vec{y} ) \hspace{1mm} (\phi^{\mathcal{O}} \circ \Lambda^{-1}) ( \Lambda ( t_n^{(1)},\vec{y} ) ) \right. \nonumber \\
  & \qquad \qquad \qquad \qquad \left. + K_2 ( \Lambda^{-1} x' ; t_n^{(2)},\vec{y} ) \hspace{1mm} (\phi^{\mathcal{O}} \circ \Lambda^{-1}) ( \Lambda( t_n^{(2)},\vec{y} ) ) \right]
\end{align}
which is an analogue of \eqref{eq:genlreconstruction_covariance}.
If one wishes, the kernels can be rewritten as
\begin{equation}
  K_i( \Lambda^{-1} x'; t_n^{(i)}, \vec{y} ) = K_i( \Lambda^{-1} x'; \Lambda^{-1} \Lambda ( t_n^{(i)},\vec{y} ) ) =: (K_i \circ \Lambda^{-1}) ( x'; \Lambda ( t_n^{(i)}, \vec{y} ) ),
\end{equation}
so that they look more like functions of only the new coordinates.
Either way, it is still consistent with \eqref{eq:genlreconstruction_covariance}.

The interesting features of the sampling, however, occur when the function space is invariant under Lorentz transformations, as in the case of functions bandlimited to $|k^2| < \Omega^2$.
Because no inertial frame is preferred, another observer in a different inertial frame $\mathcal{O}'$ should be able to employ the same reconstruction formula in their frame.
By this, we mean that using
\begin{equation}
  S_i' = \sum_{n \in \mathbb{Z}} \int d^N\vec{x} \left. |t_n^{(i)},\vec{x})_{\mathcal{O}'} \right. {}_{\mathcal{O}'} (t_n^{(i)},\vec{x}|
\end{equation}
and
\begin{equation}
  H_i' = \int \frac{d^{N+1}k}{(2\pi)^{N+1}} H_i(k) \left. |k)_{\mathcal{O}'} \right. {}_{\mathcal{O}'} (k|,
\end{equation}
then we should have $|\phi) = ( H_1' S_1' + H_2' S_2' ) |\phi)$.
Note that $t_n^{(i)}$ take the same numerical values as in the $\mathcal{O}$ frame, but now they correspond to values of the time coordinate in the $\mathcal{O}'$ frame.
Similarly, $H_i(k)$ are the same functions of $k$ as before, but now are functions of $k$ in the $\mathcal{O}'$ frame.
If frames $\mathcal{O}$ and $\mathcal{O}'$ are related by the Lorentz transformation $|x)_{\mathcal{O}'} = U(\Lambda) |x)_{\mathcal{O}}$, then $H_i' = U(\Lambda) H_i U(\Lambda)^\dagger$ and $S_i' = U(\Lambda) S_i U(\Lambda)^\dagger$.
Therefore, $|\phi) = ( H_1' S_1' + H_2' S_2' ) |\phi)$ holds since $U(\Lambda)^\dagger |\phi) = ( H_1 S_1 + H_2 S_2 ) U(\Lambda)^\dagger |\phi)$ holds due to the symmetry of the bandlimited space.
We can express $|\phi) = ( H_1' S_1' + H_2' S_2' ) |\phi)$ more concretely as:
\begin{equation}
  \phi^{\mathcal{O}'}(x) = \sum_{n \in \mathbb{Z}} \int d^N\vec{y} \left[ K_1 (x; t_n^{(1)},\vec{y}) \hspace{1mm} \phi^{\mathcal{O}'} ( t_n^{(1)},\vec{y} ) + K_2 (x; t_n^{(2)},\vec{y}) \hspace{1mm} \phi^{\mathcal{O}'} (t_n^{(2)},\vec{y}) \right],
\end{equation}
where we note that ${}_{\mathcal{O}'} (x| H_i' |y)_{\mathcal{O}'} = {}_{\mathcal{O}} (x| H_i |y)_{\mathcal{O}} \equiv K_i(x;y)$, so the kernels take the same form in this frame as in the $\mathcal{O}$ frame, which we noted in \eqref{eq:genlreconstruction_otherframes}.
Abstractly, this is equivalent to $|\phi) = U(\Lambda) ( H_1 S_1 + H_2 S_2 ) U(\Lambda)^\dagger |\phi)$, which can be written more concretely as:
\begin{equation}
  \phi^{\mathcal{O}}(x) = \sum_{n \in \mathbb{Z}} \int d^N\vec{y} \left[ K_1( \Lambda^{-1} x; t_n^{(1)},\vec{y} ) \hspace{1mm} \phi^{\mathcal{O}}( \Lambda (t_n^{(1)},\vec{y}) ) + K_2( \Lambda^{-1} x; t_n^{(2)},\vec{y} ) \hspace{1mm} \phi^{\mathcal{O}}( \Lambda (t_n^{(2)},\vec{y}) ) \right],
\end{equation}
which is an analogue of \eqref{eq:genlreconstruction_family}.
Therefore, we see that the general discussion of the previous subsection can also be applied to the Minkowski temporal sampling.
We can therefore conclude that Minkowski-bandlimited functions can be reconstructed from either a discrete set of constant-time hypersurfaces, or from a set of spacelike hypersurfaces consisting of the image of these constant-time hypersurfaces under an arbitrary Lorentz transformation.
This is the manner in which the symmetry of the function space manifests itself in the sampling theory.

\section{Conclusion}

In this paper, we have provided a thorough derivation of the temporal sampling formula for bandlimited functions on Minkowski spacetime.
We also showed how, in general, spacetime symmetries manifest themselves in sampling theory through the existence of a family of sampling lattices related by the symmetry transformations, and none of these lattices are preferred.

By examining the details of the development of the Minkowski temporal sampling formula, we were able to identify some features of the reconstruction.
One of these features was that the reconstruction kernels used to recover the Minkowski-bandlimited functions from their samples are not unique, even for the same set of discrete points in time.
We made particular choices in order to obtain an explicit solution, but fundamentally none of the choices are preferred.
If these sampling formulas are used represent quantum fields in terms of their samples in a path integral, might this arbitrariness have some physical significance?
Also, clearly the temporal sample density we chose could not be decreased with the method we employed.
However, the arbitrariness in the reconstruction kernels occurs because of gaps between the copies in Figure~\ref{fig:Minkowski_copies} (inside and outside of the bandlimited region).
It is possible that another method for developing a reconstruction formula may be able to pack these copies more tightly, and hence remove some of this arbitrariness.

Another observation we made was that the shift between the two equidistantly-spaced temporal lattices must be very finely tuned so that the solutions for the filters in various regions of Fourier space do not diverge.
As we mentioned above, this may produce an instability in the reconstruction and should be investigated.
This could be important to consider if using a sample representation for bandlimited fields in a path integral.
It could also be significant in scenarios where samples are lost or inaccessible, which would occur naturally, e.g., for uniformly accelerated observers.

There are also a number of other directions of inquiry which could stem from our analyses here.
For instance, we demonstrated that if one has a family of reconstruction formulas of the form \eqref{eq:genlreconstruction_family} on some function space, then this space can always be extended to one which is invariant under a symmetry and where these reconstruction formulas hold.
However, here one assumes that the reconstruction kernels used for each lattice are related in a very particular manner.
One may consider whether this hypothesis can be weakened.
For example, we know that since there is generically freedom in the choice of kernels, it is possible to have a set of kernels different than those in \eqref{eq:genlreconstruction_family}.
This suggests that one may be able to make a milder assumption regarding the kernels in order to demonstrate that a family of sampling lattices implies a symmetry for the function space.
It is also possible that the behaviour of the sampling theory under general diffeomorphisms may play an interesting role here.

Of course, there is also the question of how sampling theory behaves under general coordinate transformations.
This may be important for progress in understanding sampling for the generally-covariant notion of bandlimitation that we briefly mentioned in Subsection~\ref{subsec:sampling_overview}.
For this generalization of sampling theory, it should be interesting to see the effects of spacetime curvature, as well as what happens in the presence of horizons and singularities.
For example, sampling theory on a black hole spacetime may provide a unique perspective on the localization of field degrees of freedom, and could be insightful for the information paradox and holographic principle.

Our discussion in Subsection~\ref{subsec:genl_considerations} may also be helpful in constructing new function spaces which have the sampling property and are invariant under some symmetry group.
In the beginning of this paper, we placed a primary importance on establishing some appropriate notion of bandlimitation in extending sampling theory to new situations.
However, some of the observations we made here may allow for a more direct construction of function spaces with these properties without first concocting different notions of bandlimitation.

We also note that some of the discussion here appears conceptually similar to statements regarding the Lorentz invariance of causal sets \cite{BombelliHensonSorkin2009,Dowker2011,DowkerSorkin2020}.
In particular, in causal set theory one has a discrete spacetime structure without picking out a preferred Lorentz frame.
One is then led to wonder whether causal sets could be related to sampling theory in some way.

In addition to further understanding the mathematics of sampling theory in spacetime, ultimately the aim is to continue to develop physical applications for this model.
Of course, one important direction is to search for futher possible observational signatures.
Another point of interest is to determine the impact that the Minkowski bandlimit has on interacting quantum field theories.
Minkowski bandlimitation can be interpreted as cutting out four-momenta which are far off-shell, and so this may have a significant effect on the structure of the divergences which occur from integrating over loop momenta in the Feynman diagrams of interacting theories.

\section*{Acknowledgments}
The author would like to thank Achim Kempf for many invaluable discussions, and for his continuing support and generosity.
The author would also like to thank an anonymous referee for helpful suggestions to improve the paper.

\bibliographystyle{ieeetr}
\bibliography{main}

\end{document}